\documentclass[12pt,preprint]{aastex}

\def\apj{{ApJ}}

\def\be{\begin{equation}}
\def\ee{\end{equation}}
\def\bea{\begin{eqnarray}}
\def\eea{\end{eqnarray}}

\usepackage{color,txfonts}
\usepackage{hyperref}
\usepackage{epstopdf}
\usepackage{graphicx}
\usepackage[FIGTOPCAP]{subfigure}
\usepackage{CJKutf8}

\begin{document}
\title{Short GRBs: opening angles, local neutron star merger rate and off-axis events for GRB/GW association}
\author{Zhi-Ping Jin$^{1,2}$, Xiang Li$^{1}$, Hao Wang$^{1,2}$, Yuan-Zhu Wang$^{1,2}$, Hao-Ning He$^{1}$, Qiang Yuan$^{1,2}$,
Fu-Wen Zhang$^{3}$, Yuan-Chuan Zou$^{4}$, Yi-Zhong Fan$^{1,2}$, and Da-Ming Wei$^{1,2}$}
\affil{
$^1$ {Key Laboratory of dark Matter and Space Astronomy, Purple Mountain Observatory, Chinese Academy of Science, Nanjing, 210008, China.}\\
$^2$ {School of Astronomy and Space Science, University of Science and Technology of China, Hefei, Anhui 230026, China.}\\
$^3$ {College of Science, Guilin University of Technology, Guilin 541004, China.}\\
$^4$ {School of Physics, Huazhong University of Science and Technology, Wuhan 430074, China.}\\
}
\email{yzfan@pmo.ac.cn (YZF) and dmwei@pmo.ac.cn (DMW)}

\begin{abstract}
The jet breaks in the afterglow lightcurves of short gamma-ray bursts (SGRBs), rarely detected so far, are crucial for estimating the half-opening angles of the ejecta ($\theta_{\rm j}$) and hence the neutron star merger rate. In this work we report the detection of jet decline behaviors in GRB 150424A and GRB 160821B and find $\theta_{\rm j}\sim 0.1$ rad. Together with five events reported before 2015 and other three ``identified" recently (GRB 050709, GRB 060614 and GRB 140903A), we have a sample consisting of nine SGRBs and one long-short GRB with reasonably estimated $\theta_{\rm j}$. In particular, three {\it Swift} bursts in the sample have redshifts $z\leq 0.2$, with which we estimate the local neutron star merger rate density {to be  $\sim 1109^{+1432}_{-657}~{\rm Gpc^{-3}~yr^{-1}}$  or $162^{+140}_{-83} {\rm Gpc^{-3}yr^{-1}}$ if the narrowly-beamed GRB 061201 is excluded}. Inspired by the typical $\theta_{\rm j}\sim 0.1$ rad found currently, we further investigate whether the off-beam GRBs (in the uniform jet model) or the off-axis events (in the structured jet model) can significantly enhance the GRB/GW association or not. For the former the enhancement is at most moderate, while for the latter the enhancement can be much greater and a high GRB/GW association probability of $\sim 10\%$ is possible. We also show that the data of GRB 160821B may contain a macronova/kilonova emission component with a temperature of $\sim 3100$ K at $\sim 3.6$ days after the burst and more data are needed to ultimately clarify.
\end{abstract}
\keywords{gamma-ray burst: individual: GRB 150424A, GRB160821B ---binaries: close---gravitational wave}

\section {Introduction}

Gamma-ray Bursts are brief, intense gamma-ray flashes that are widely believed to be powered by the dying massive stars (also called as collapsars) or the merging of the compact binaries involving at least one neutron star \citep[see][for reviews]{Piran2004,Kumar2015}. Bright supernova (SN) emissions have been detected in the late afterglows of some low-redshift long duration GRBs (LGRBs, i.e., GRBs with durations longer than 2 seconds), established their collapsar origin \citep{Woosley2006}. The neutron star merger model is more likely relevant to short GRBs (SGRBs), the events with durations shorter than 2 seconds \citep{Kouveliotou1993,Eichler1989}.
The direct evidence for the merger origin of SGRBs was absent until August 16, 2017 due to the non-detection of SGRB/GW association.
There was, however, some indirect evidence for SGRBs originating from compact binaries,  including the location of
SGRBs in elliptical galaxies, non-detection of associated SN, large galaxy offsets, weak spatial correlation
of SGRBs and star formation regions within their host
galaxies \citep[see][for reviews]{Nakar2007,Berger2014}, and in particular the identification of the so-called Li-Paczynski macronovae in GRB 130603B \citep{Tanvir2013,Berger2013}, GRB 060614 \citep{Yang2015,Jin2015} and GRB 050709 \citep{Jin2016}. The Li-Paczynski macronova (also called a kilonova)  is a new
kind of near-infrared/optical transient, powered by the radioactive
decay of {\it r}-process material, which is synthesized in the ejecta
launched during the merger event \citep[e.g.,][]{Li1998,Kulkarni2005,Metzger2010,Barnes2013,Tanaka2013,Hotokezaka2013}. On the other hand, the mergers of
compact binaries consisting of neutron stars (NSs) and/or black holes (BHs) are promising sources of gravitational waves \citep[GW;][]{Clark1977}.
Therefore, SGRBs are expected to be one of the most important electromagnetic counterparts of the GW events \citep[e.g.,][]{Eichler1989,Kochanek1993,LiX2016ApJ,Baiotti2017,Paschalidis2017}. The successful detection of GW170817/GRB170817A/AT2017gfo \citep{Goldstein2017,LVC2017,Coulter2017,Pian2017}, {\it reported during the revision of this work,} directly confirm all the above opinions/speculations.

The SGRB data had been widely used to estimate the rate of neutron star mergers and hence the detection prospect of the GW detectors \citep[e.g.,][]{Guetta2005,Nakar2007,Abadie2010,Coward2012,Fong2013,Fong2014,LiX2017}. Such an approach is still necessary
after the successful detection of GW170817. This is because, though the mergers of double neutron stars (neutron star-black hole binaries) can be directly measured by advanced LIGO/Virgo, the horizon distances of detecting such events by the second generation detectors are limited to be $z<0.1~(0.2)$ even under the optimistic situation (for example, the successful detection of the electromagnetic counterparts). Therefore, at relatively higher redshifts, the SGRB data are valuable in estimating the neutron star merger rate. At low redshifts, the comparison of the geometry-corrected SGRB rate with the directly measured neutron star merger rate can be used to evaluate the probability of launching relativistic outflow in these mergers.
For such purposes, the half-opening angles of the SGRB ejecta are crucial.
In principle, the half-opening angle of GRB ejecta can be reasonably estimated with the measured jet break in the afterglow light curve \citep{Rhoads1999,Sari1999}. Such jet breaks are unfortunately only rarely detected for SGRBs. As summarized in \citet{Fong2015}, previously there were just four SGRBs, including GRB 051221A, GRB 090426A, GRB 111020A and GRB 130603B, with reliable jet break measurements and hence the half-opening angle estimates (GRB 061201 with a sharp jet break displaying in the X-ray data \citep{Stratta2006} was however ignored in such a summary). The deep X-ray measurement of GRB 140903A reveals a jet break at $\sim 1$ day after the burst \citep{Troja2016a}, suggesting a $\theta_{\rm j}\sim 0.05-0.1$ rad \citep{ZhangS2017}. The sample increases a bit further if we also include the jet break displaying in the so-called long-short GRB (lsGRB) 060614 \citep{Xu2009,Jin2015} and the quick decline behavior needed in modeling of short GRB 050709 \citep{Jin2016}. Even so, the ``enlarged" sample consists of just $<7\%$ SGRBs and it is unclear whether the majority of the SGRBs are beamed or not. One speculation is that the jet breaks usually take place very late and deep follow-up observations are needed. The search for macronova/kilonova components in late time afterglow emission of SGRBs, motivated by the signal detected in GRB 130603B, provided us such a chance. In this work we analyzed the Hubble Space Telescope (HST) data of GRB 150424A \citep{Tanvir2015} and GRB 160821B \citep{Troja2016b} to search for the jet breaks as well as the macronova signals.

This work is organized as the following. In section 2, we describe our data analysis of GRB 150424A and GRB 160821B, and identify the jet breaks (or more exactly, the post-jet-break decline behavior). We also examine whether there is evidence for a macronova emission component in GRB 160821B, and estimate the neutron star merger rate density in the local universe. In section 3, inspired by the narrow jets found in the current sample, we investigate whether the off-beam GRBs (in the uniform jet model, the line of sight is outside of the ejecta) or the off-axis events (in the structured jet model,  the line of sight is outside of the narrow energetic core of the ejecta) can significantly contribute to the GRB/GW association or not. We summarize our results with some discussions in section 4.

\section{Data analysis of GRB 150424A and GRB 160821B, jet breaks of SGRBs/lsGRB, and the local neutron star merger rate}

\subsection{Observations and data analysis of GRB 150424A and GRB 160821B}

SGRB 150424A was detected by the Burst Alert Telescope (BAT) onboard {\it Swift} satellite at 07:42:57 (UT) on April 24, 2015 \citep{Beardmore2015}.
The BAT light curve is composed by a bright multi-peaked short episode (start from $T-0.05$ to $T+0.5$ second) and a very weak extended emission (lasted to about $T+100$ seconds) \citep{Barthelmy2015}. Konus-Wind also detected a multi-peak short burst with total duration of $\sim0.4$ second \citep{Golenetskii2015}.
At 1.6 hours after the trigger, Keck telescope observed the field and found the optical afterglow \citep{Perley2015}.
\citet{Marshall2015} checked the earlier {\it Swift} UV optical telescope (UVOT) observations and found the afterglow, too.
HST visited the burst site for three epochs within a month (PI: Nial Tanvir, HST proposal ID: 13830).
Recently, \citet{Knust2017} reported the multi-wavelength observation data of GRB 150424A and interpreted the central engine as a magnetar.
In this work we focus on the late observations ($>$0.5 day) to search for possible jet break and the macronova signal.

SGRB 160821B was detected by the {\it Swift} BAT at 22:29:13 (UT) on August 21, 2016 \citep{Siegel2016}.
Its duration is T$_{90}=0.48\pm0.07$ second.
Following the BAT trigger, the X-ray telescope (XRT) and UVOT slew to the burst site immediately and started the observation in 66 and 76 seconds, respectively.
The XRT found the afterglow at R.A.=18:39:54.71 Dec.=62:23:31.3 (J2000) with an uncertainty of 2.5 arcseconds \citep{Siegel2016}.
Based on the observation started at 0.548 hour after the burst, the Nordic Optical Telescope (NOT) firstly reported the detection of the optical afterglow at R.A.=18:39:54.56 Dec.=62:23:30.5 (J2000) with an uncertainty of 0.2 arcsecond, and a host galaxy candidate which lies at about 5.5" away \citep{Xu2016}. Later, the William Herschel Telescope spectral observations of the host galaxy claimed a redshift of $z=0.16$ based on the H$\alpha$, H$\beta$ and O${\rm III} 4959/5007\AA$
lines. The unambiguous classification as SGRB and the plausible low redshift make it an ideal candidate to search for a macronova. HST visited the burst site for three epochs within a month (PI: Nial Tanvir, HST proposal ID: 14237).

Now the HST data of both GRB 150424A and GRB 160821B are publicly-available and we have downloaded these data from the Barbara A. Mikulski Archive for Space Telescopes (MAST: http://archive.stsci.edu/).

\begin{table}
\begin{center}
\label{tab:GRB150424A}
\title{}Table 1. Observations of GRB150424A.\\
\begin{tabular}{lllll} \hline \hline
Time	& Exposure	& Instrument	& Filter	& Magnitude$^{a}$\\
(days)	& (seconds)	& 	& 	& (AB) \\ \hline \hline
177.44649	& 2496	& HST+WFC3	& $F475W$  & ($>27.4$)$^{b}$\\
6.63693	& 1800	& HST+WFC3	& $F606W$  & $26.03\pm0.06$\\
9.22345	& 1800	& HST+WFC3	& $F606W$  & $26.98\pm0.14$ \\
13.86864	& 1800	& HST+WFC3	& $F606W$  & $28.03\pm0.37$\\
178.47184	& 1800	& HST+WFC3	& $F606W$  & ($>28.2$)$^{b}$\\
177.51281	& 2496	& HST+WFC3	& $F814W$  & ($>26.9$)$^{b}$\\
177.61249	& 5395	& HST+WFC3	& $F105W$  & ($26.20\pm0.11$)$^{b}$\\
6.67527	& 1498	& HST+WFC3	& $F125W$  & $25.25\pm0.08$\\
9.26352	& 1498	& HST+WFC3	& $F125W$  & $26.32\pm0.21$\\
13.92497	& 1498	& HST+WFC3	& $F125W$  & $27.08\pm0.42$\\
178.51300	& 1498	& HST+WFC3	& $F125W$  & ($26.22\pm0.19$)$^{b}$\\
6.70905	& 1498	& HST+WFC3	& $F160W$  & $25.08\pm0.07$\\
9.29904	& 1498	& HST+WFC3	& $F160W$  & $25.83\pm0.14$\\
13.96139	& 1498	& HST+WFC3	& $F160W$  & $27.10\pm0.44$\\
178.56572	& 1498	& HST+WFC3	& $F160W$  & ($25.67\pm0.13$)$^{b}$\\
\hline
\end{tabular}
\end{center}
\textbf{Notes.}
a. These values have not been corrected for the Galactic extinction of $A_{\rm V}=0.06$ mag \citep{Schlafly2011} \\
b. Quote to the underlying source S1, which lies at about 0.2 second southeast.\\
\end{table}

\begin{table}
\label{tab:GRB160821B}
\begin{center}

\title{}Table 2. Observations of GRB 160821B\\

\begin{tabular}{llllll} \hline \hline
Time$^{a}$	& Exposure	& Instrument	& Filter	& Magnitude$^{b}$\\
(days)	& (seconds)	& 	& 	& (AB)\\ \hline \hline

3.64375	& 2484	& HST+WFC3	& $F606W$	& $26.00\pm0.04$\\
10.39618	& 2484	& HST+WFC3	& $F606W$	& $28.0\pm0.3$\\
23.16018	& 1350	& HST+WFC3	& $F606W$	& $>27.9^{b}$\\
103.40406	& 2484	& HST+WFC3	& $F606W$	& $-^{c}$\\

3.77481	& 2397	& HST+WFC3	& $F110W	$	& $24.78\pm0.03$\\
10.52723	& 2397	& HST+WFC3	& $F110W	$	& $26.7\pm0.2$\\
23.18086	& 1498	& HST+WFC3	& $F110W	$	& $>28.0^{b}$\\
99.26591	& 5395	& HST+WFC3	& $F110W	$	&  $-^{c}$\\

3.70861	& 2397	& HST+WFC3	& $F160W$	& $24.50\pm0.04$\\
10.46105	& 2397	& HST+WFC3	& $F160W$	& $26.9\pm0.3$\\
23.23040	& 2098	& HST+WFC3	& $F160W$	& $>27.0$\\

\hline
\end{tabular}
\end{center}
a. These values have not been corrected for the Galactic extinction of $A_{\rm V}=0.04$ mag \citep{Schlafly2011}.\\
b. Images have been combined with later exposures as references.\\
c. Images have been combined to earlier exposures as references.\\
\end{table}

\begin{figure}[ht]
\begin{center}
\includegraphics[width=0.6\textwidth]{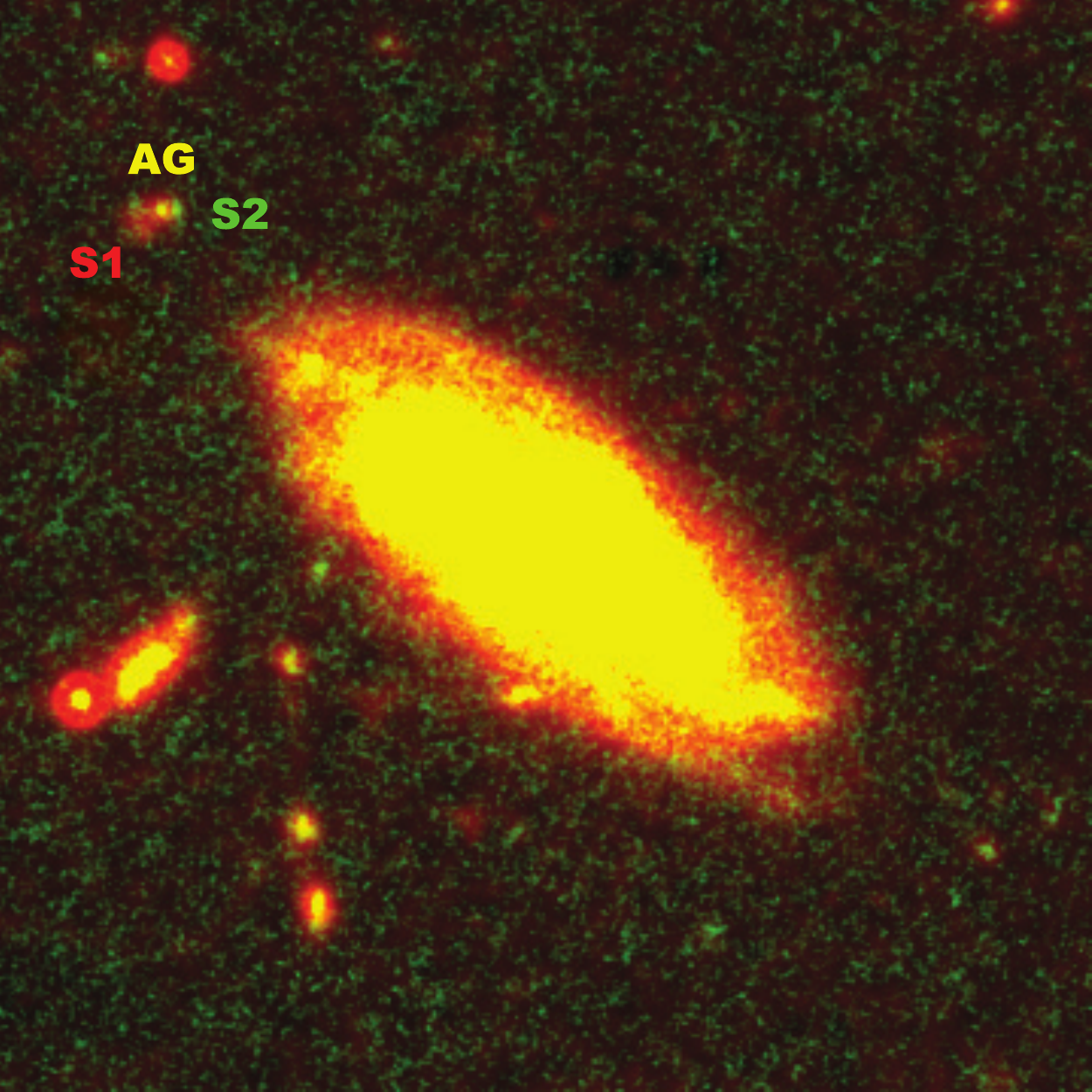}
\end{center}
\caption{False color image of the field of GRB 150424A, which has been converted to north up and east left.
Here all HST WFC3 UVIS (including 2496 seconds in $F435W$,
7200 seconds in $F606W$ and 2496 seconds in $F814W$) signals appear green and all HST WFC3 IR (including 5395 seconds in $F105W$,
5991 seconds in $F125W$ and 5991 seconds in $F160W$) appear red. The afterglow detected in both UV and IR images are marked as AG.
There are a red source in the south east only detected in IR images (marked as S1) and a blue source in the west only detected in UV images (marked as S2).
}
\label{fig:GRB150424A_host}
\end{figure}

\begin{figure}[ht]
\begin{center}
\subfigure[$F475W$] {\includegraphics[width=0.15\textwidth]{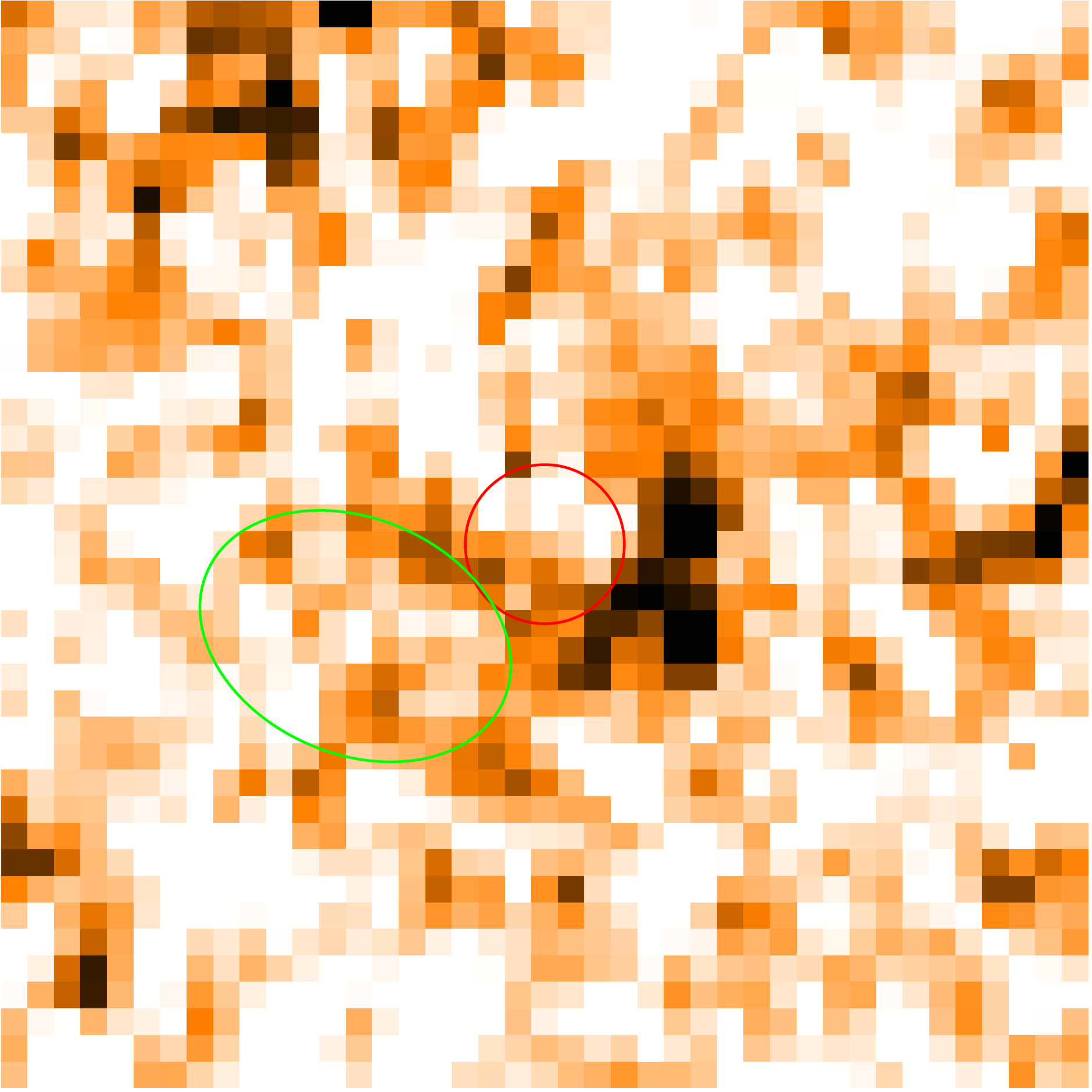}}
\subfigure[$F606W$] {\includegraphics[width=0.15\textwidth]{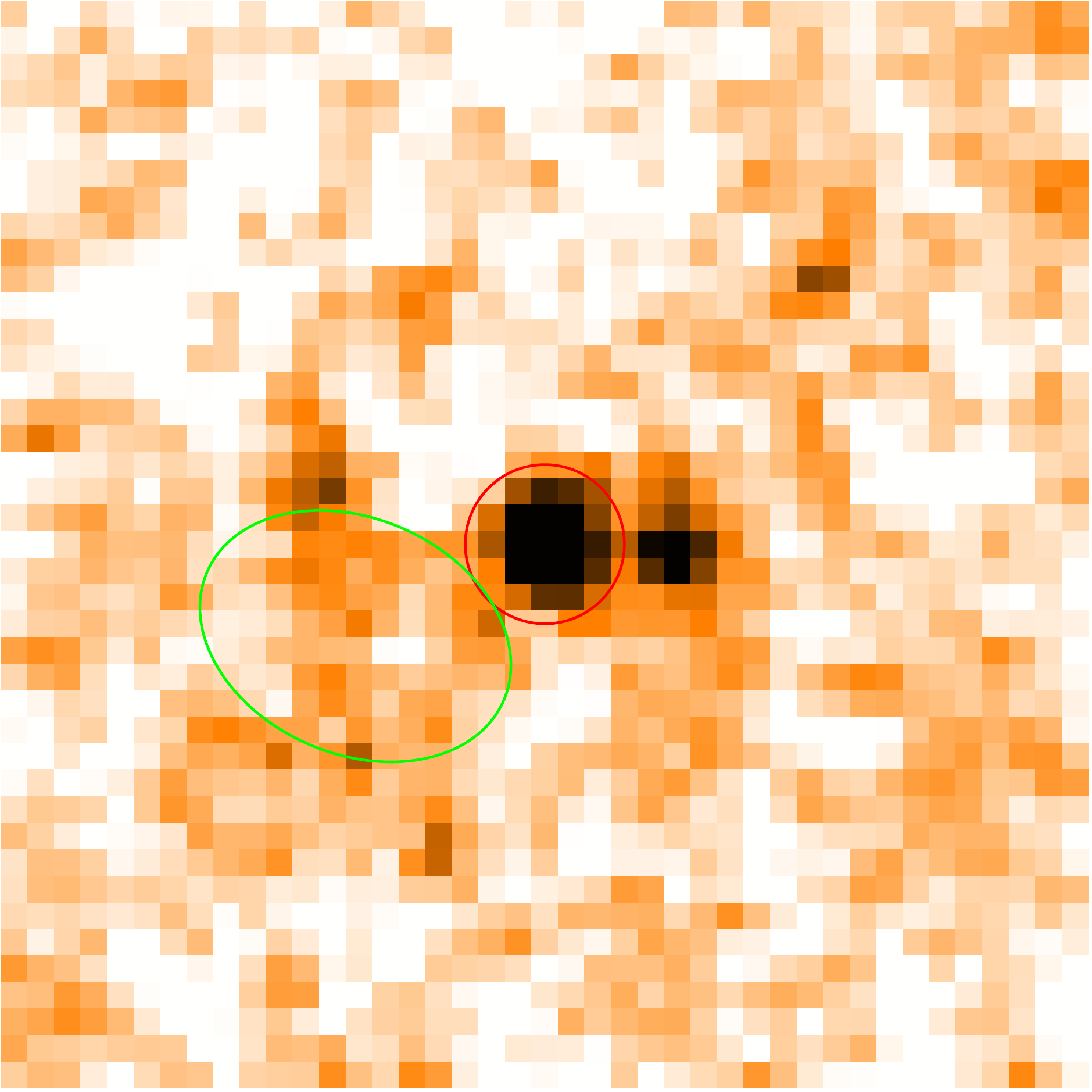}}
\subfigure[$F814W$] {\includegraphics[width=0.15\textwidth]{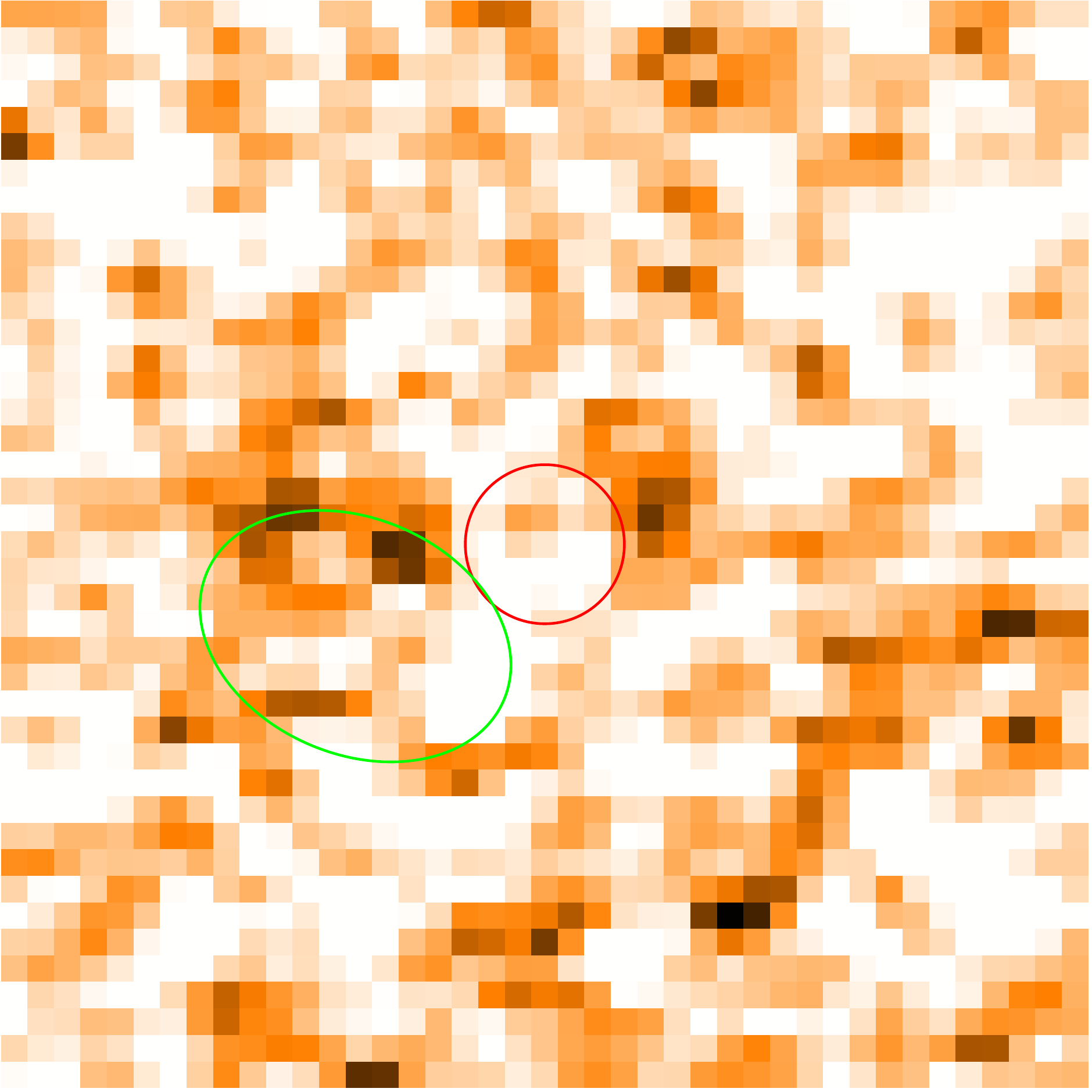}}
\subfigure[$F105W$] {\includegraphics[width=0.15\textwidth]{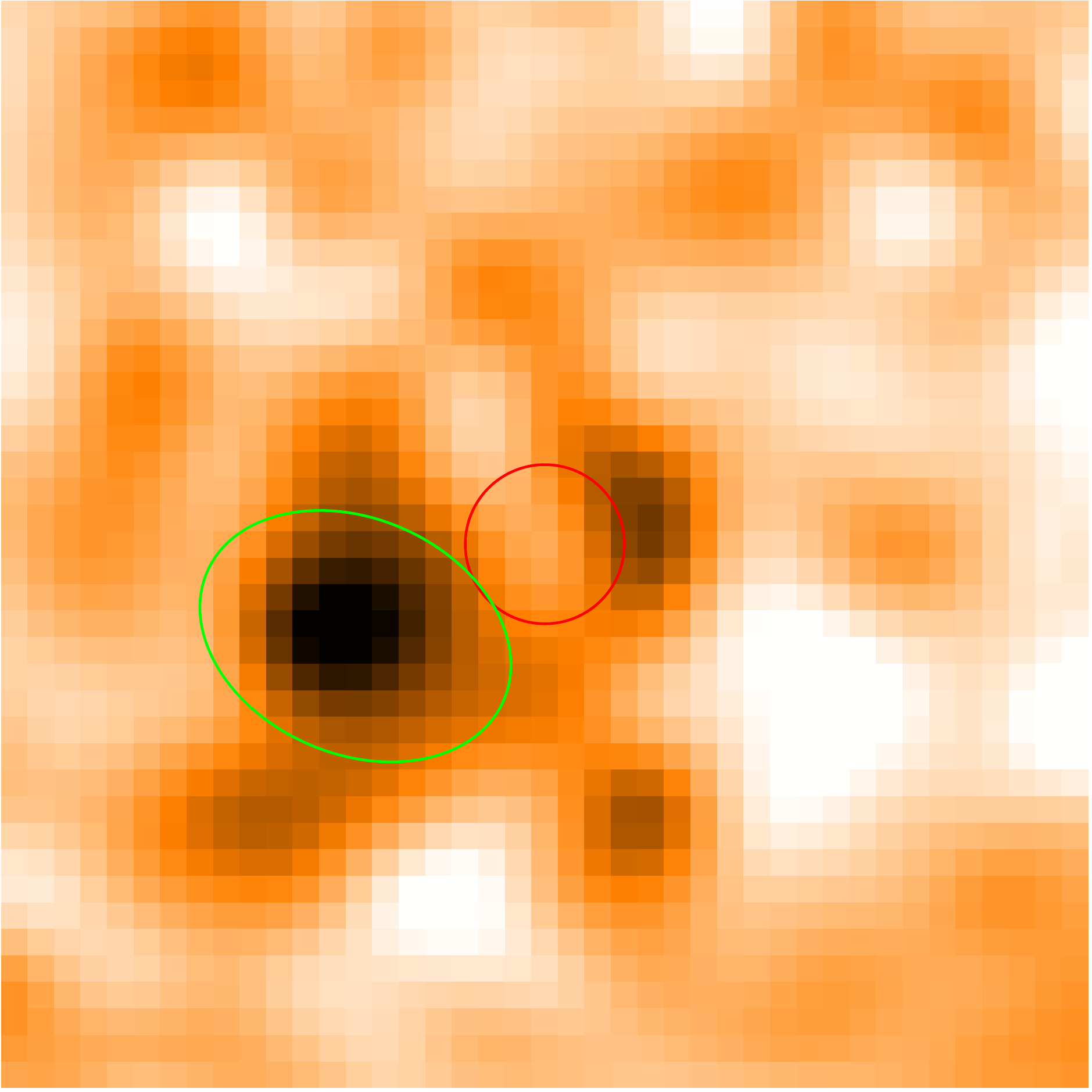}}
\subfigure[$F125W$] {\includegraphics[width=0.15\textwidth]{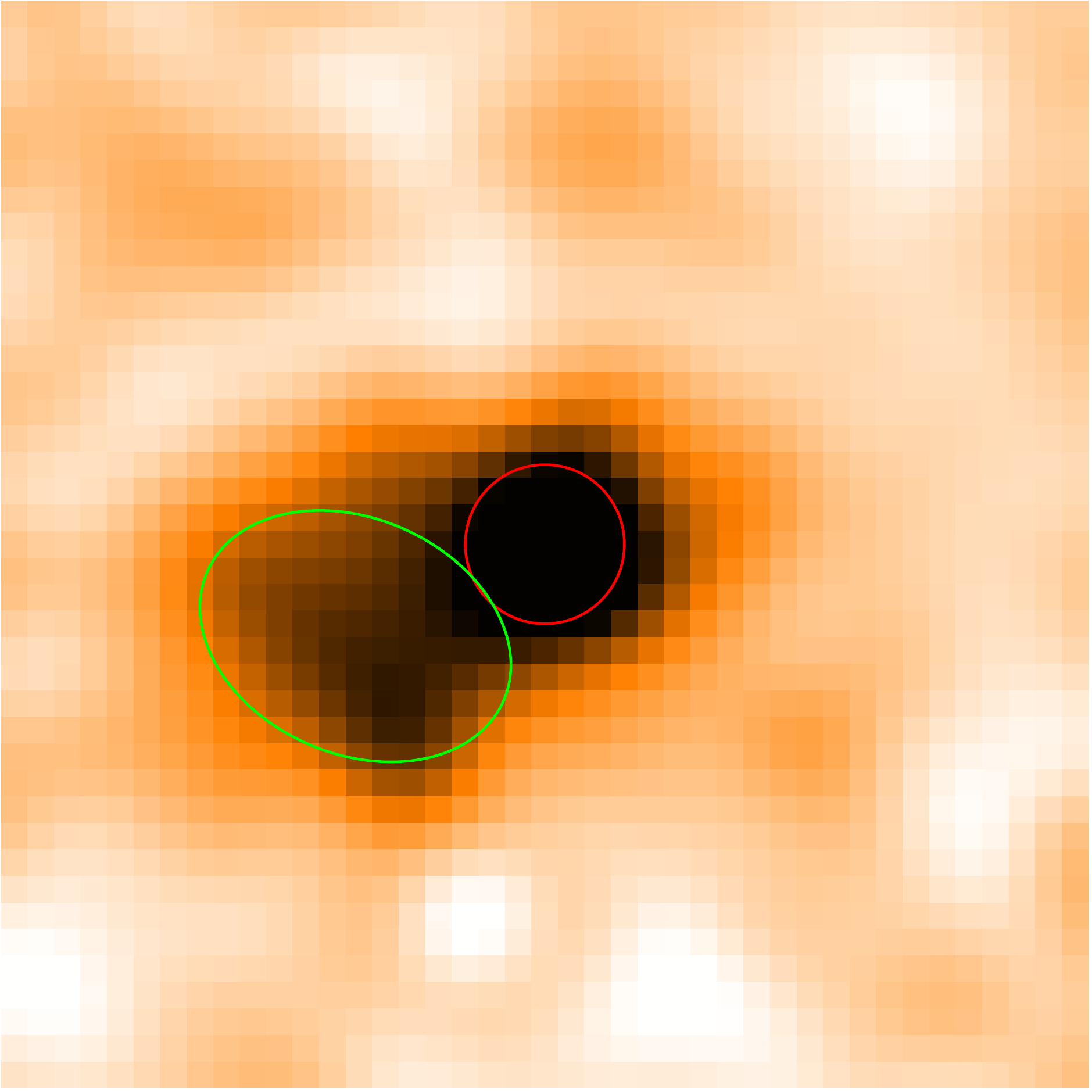}}
\subfigure[$F160W$] {\includegraphics[width=0.15\textwidth]{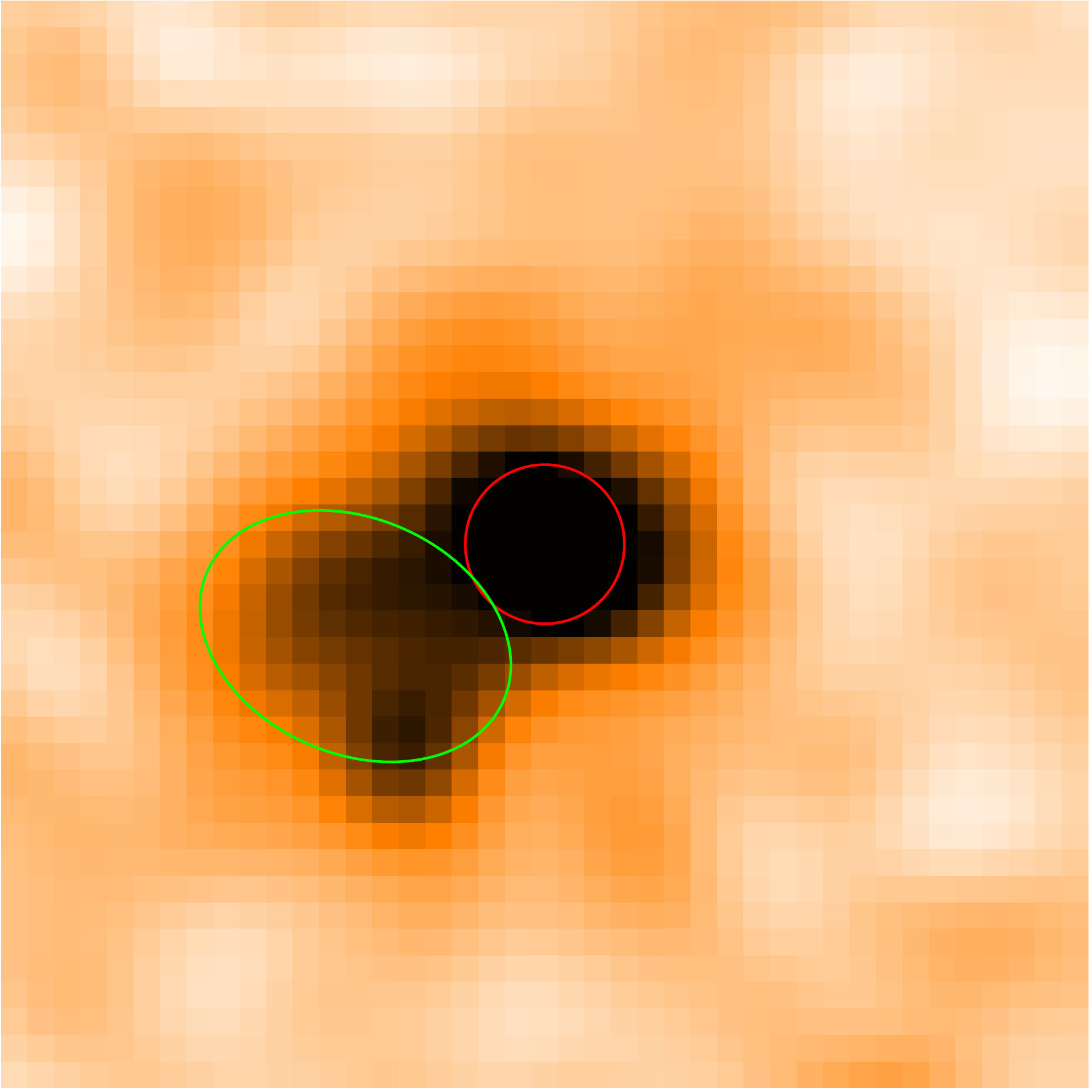}}
\end{center}
\caption{The HST observations of GRB 150424A.
Each image shows the combined data of all observations in the given filter.
The red circle is the position of the afterglow and the green circle is the position of S1.}
\label{fig:GRB150424_S1}
\end{figure}

GRB 150424A was initially found near a bright spiral galaxy $R\sim20$ mag \citep{Perley2015} at a redshift of $z=0.30$ \citep{Castro-Tirado2015}.
The later deep HST observations, however, found that the field of GRB 150424A is very complicated \citep{Tanvir2015}.
There is an extended source to the south-east of afterglow (see Fig.\ref{fig:GRB150424A_host}, marked as S1),
which is a candidate of the host galaxy of GRB 150424A, too.
It has been detected in HST F105W, F125W and F160W bands, but not in HST F475W, F606W and F814W bands, see Fig.\ref{fig:GRB150424_S1}.
If the break between F105W and F814W is due to the 4000${\rm \AA}$ break, then the redshift is $z\sim1.2$.
While the absence of strong evidence for a Lyman break in the {\it Swift} UVOT afterglow data \citep{Marshall2015,Knust2017}
indicates a redshift of $z\leq 1.0$.
Considering these uncertainties, we suggest that $z=0.3$ is favored while $z\sim 1$ is still possible.
 The HST data of GRB 150424A have been analyzed by \citet{Knust2017} via a photometry with a { 0.4"} aperture. We have carried out an independent analysis of HST data in a slightly different method. We have adopted the image subtraction technique, taking all the observations at $t>20$ days as the reference images, to remove the underlying sources (i.e., S1 and S2) from the afterglow emission. The afterglow is very weak in the second HST observation epoch. We have taken { a 3 pixels aperture} for the photometry, and then corrected to an infinite aperture. The results are reported in Tab.1. Essentially, our results are in agreement with \citet{Knust2017}, except that their photometry
included the underlying sources S1 and S2, while we have removed them by image subtraction,
hence our afterglow emission are weaker than \citet{Knust2017}.

\begin{figure}[ht]
\begin{center}
\subfigure[F606W e1] {\includegraphics[width=0.15\textwidth]{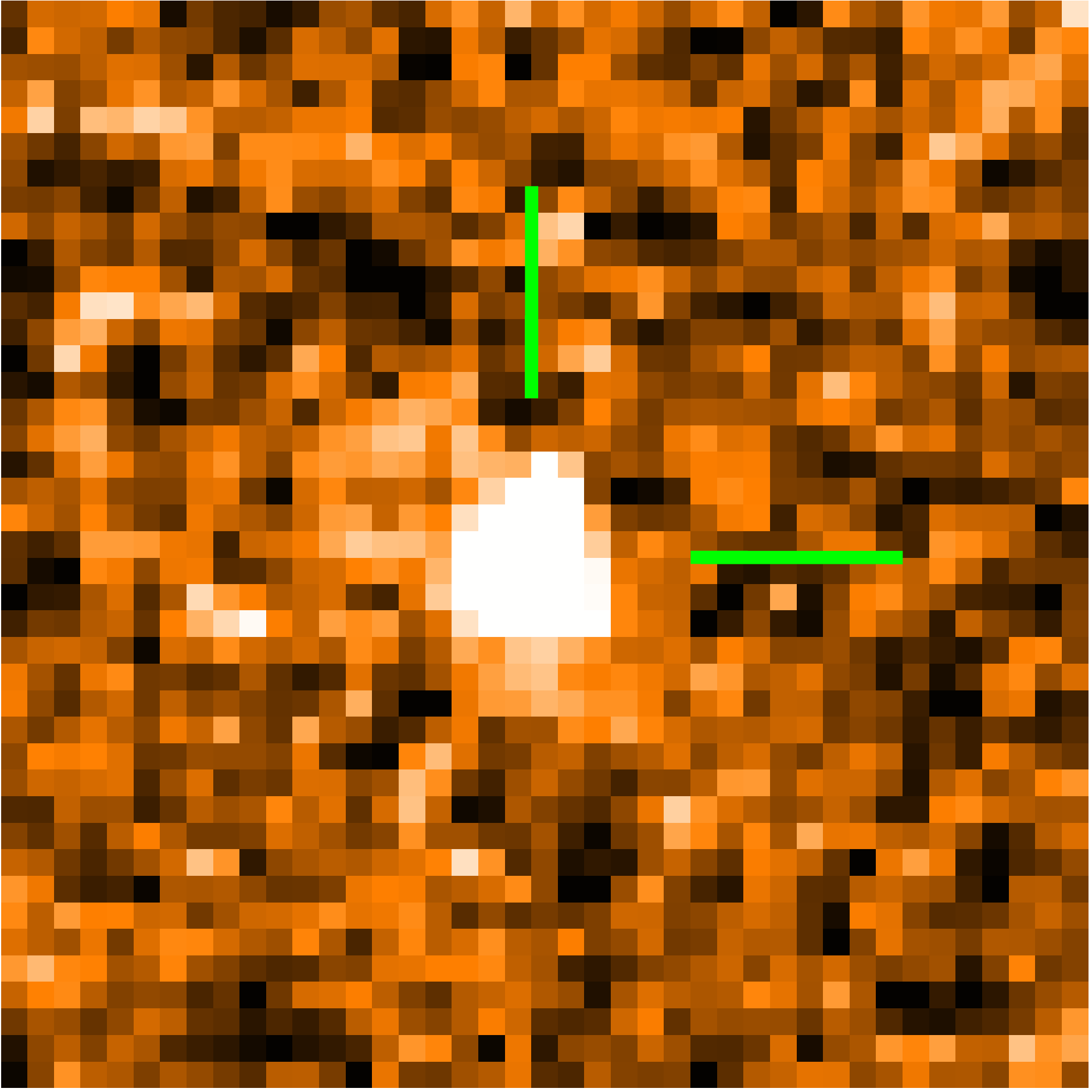}}
\subfigure[F606W e2] {\includegraphics[width=0.15\textwidth]{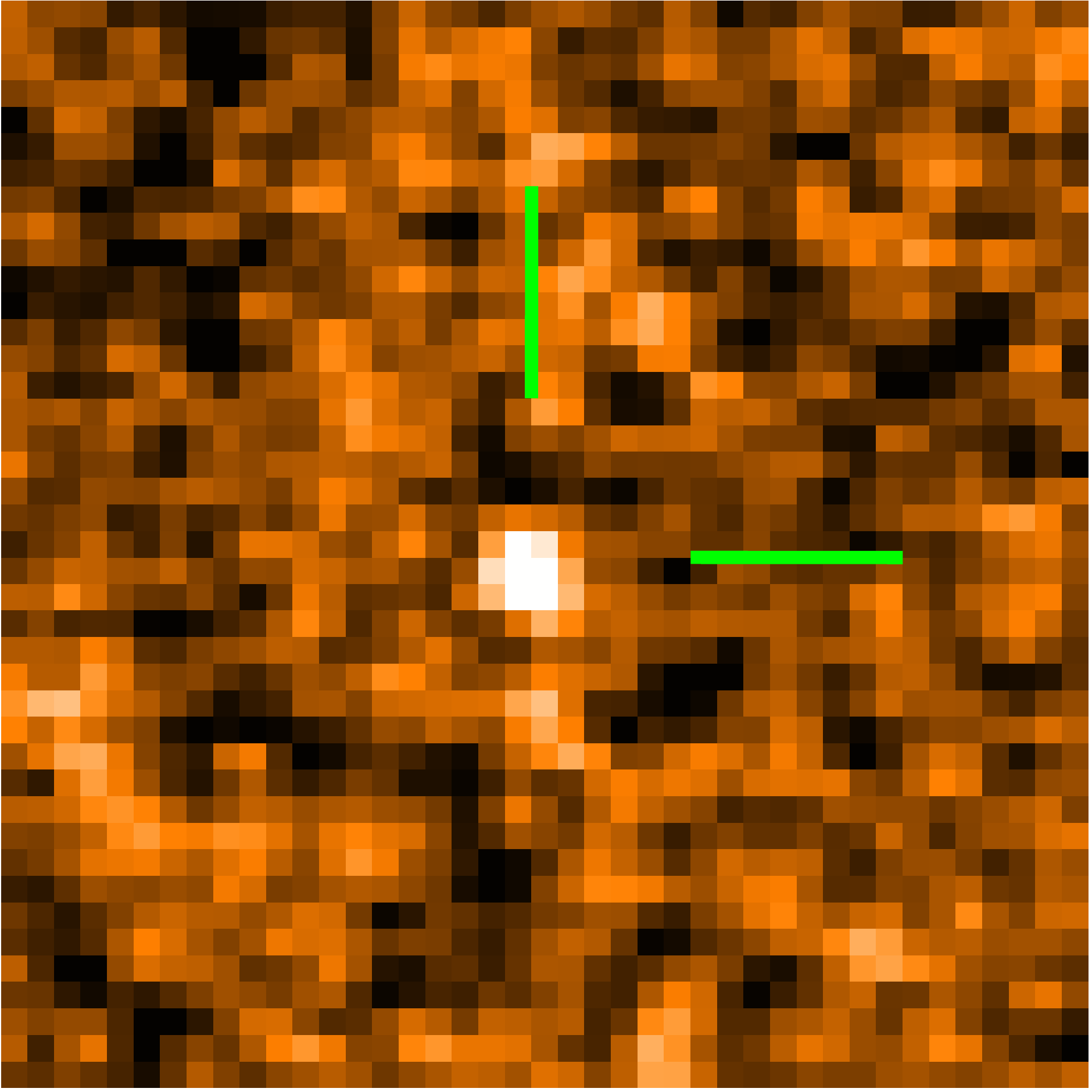}}
\subfigure[F606W ref] {\includegraphics[width=0.15\textwidth]{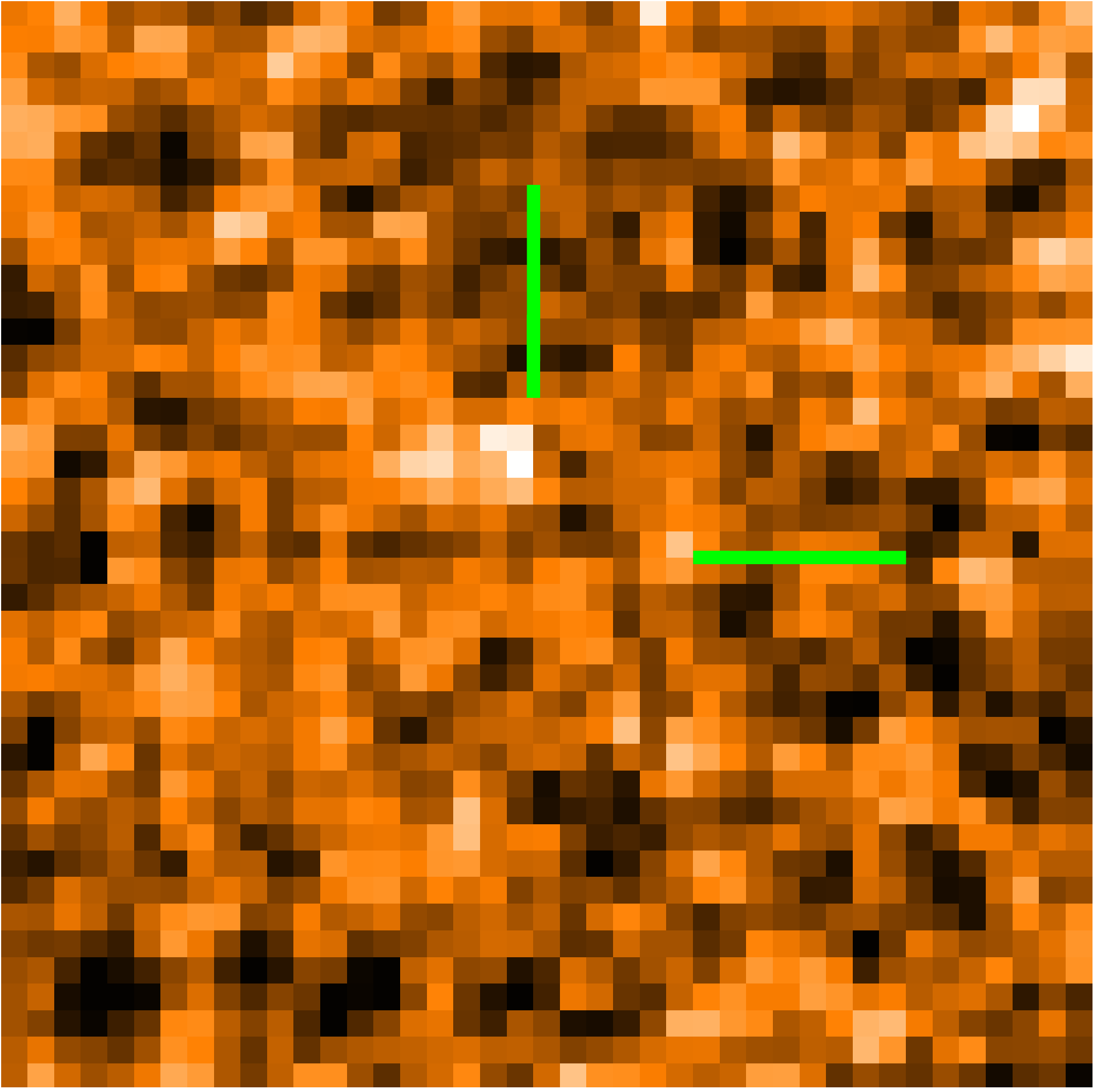}}\\
\subfigure[F110W e1] {\includegraphics[width=0.15\textwidth]{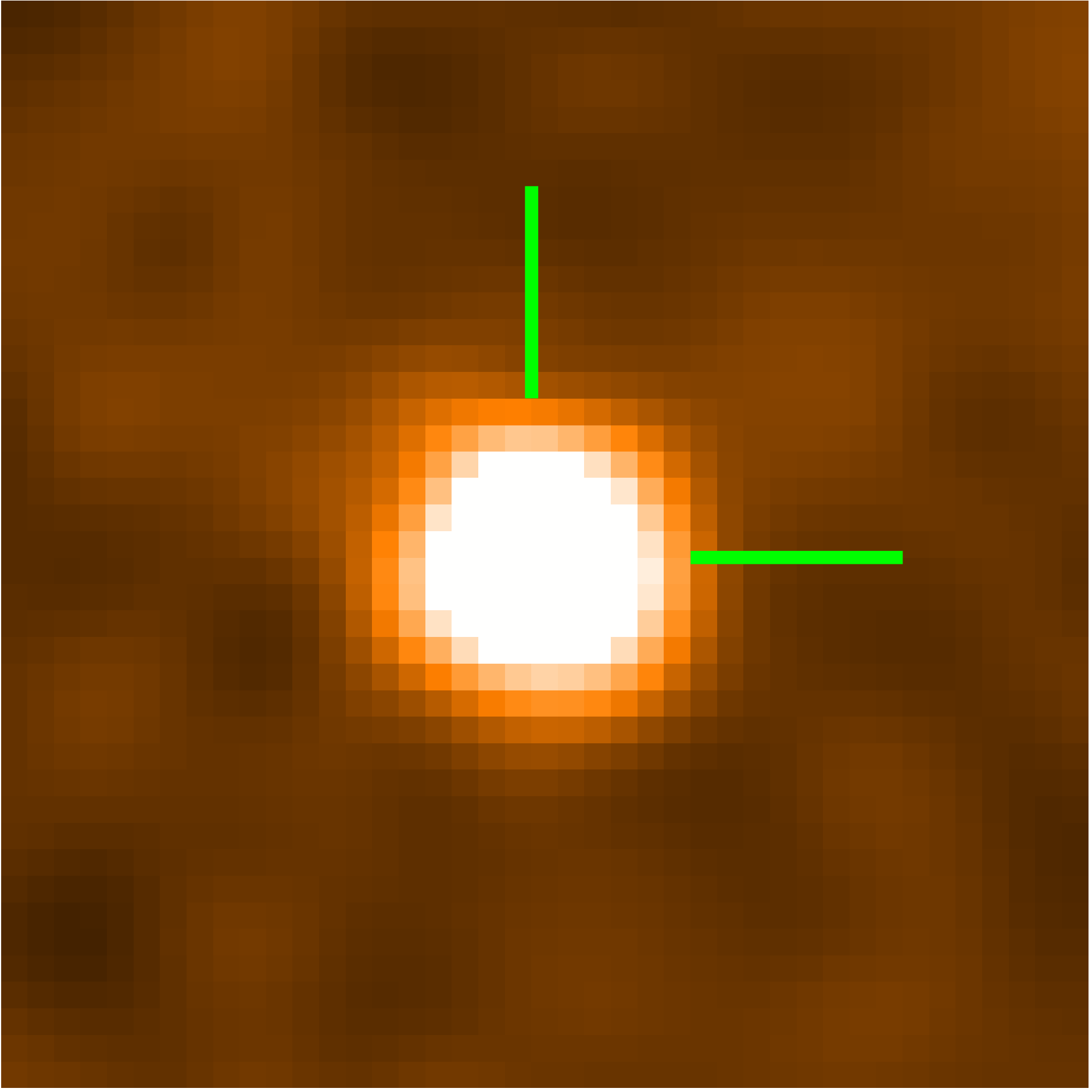}}
\subfigure[F110W e2] {\includegraphics[width=0.15\textwidth]{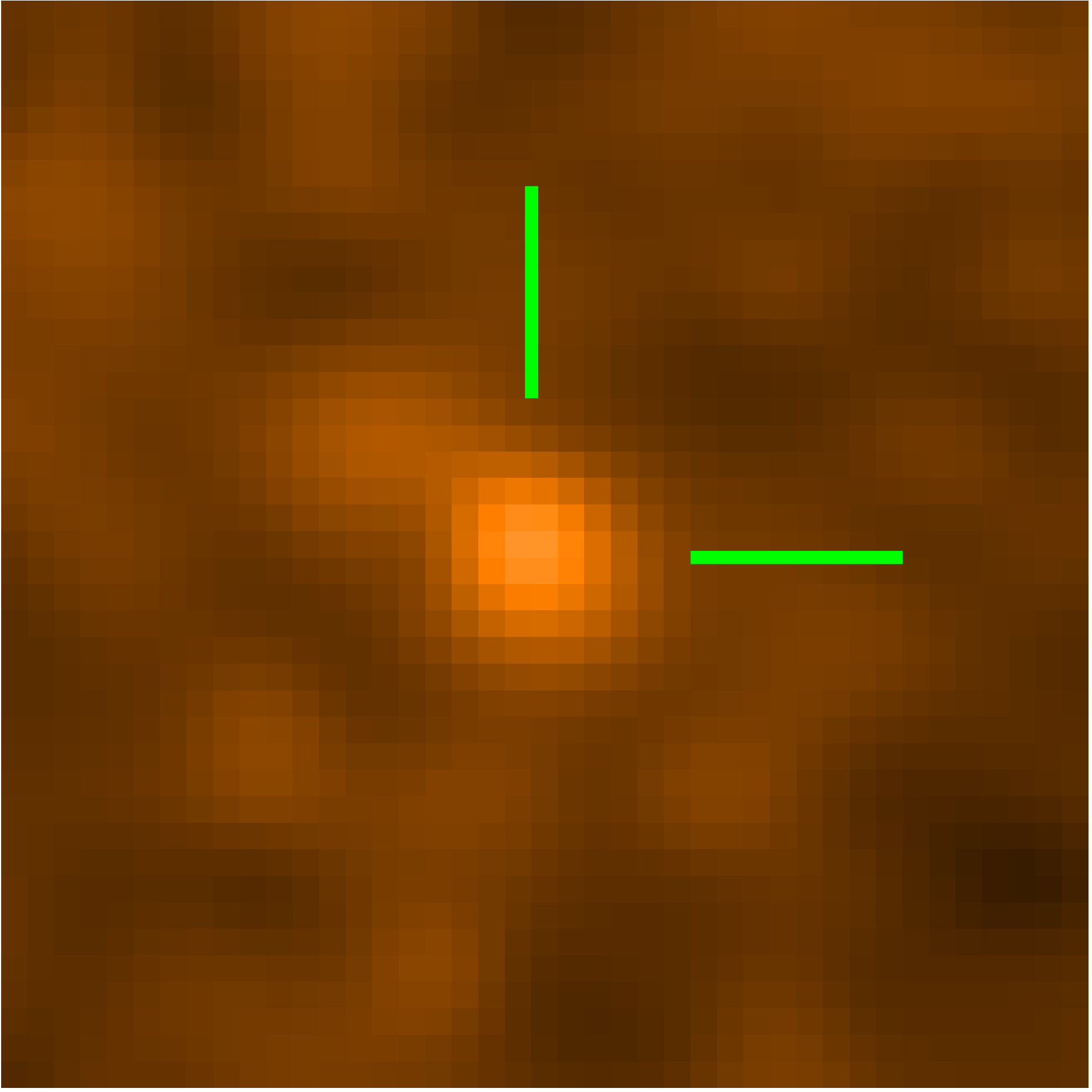}}
\subfigure[F110W ref] {\includegraphics[width=0.15\textwidth]{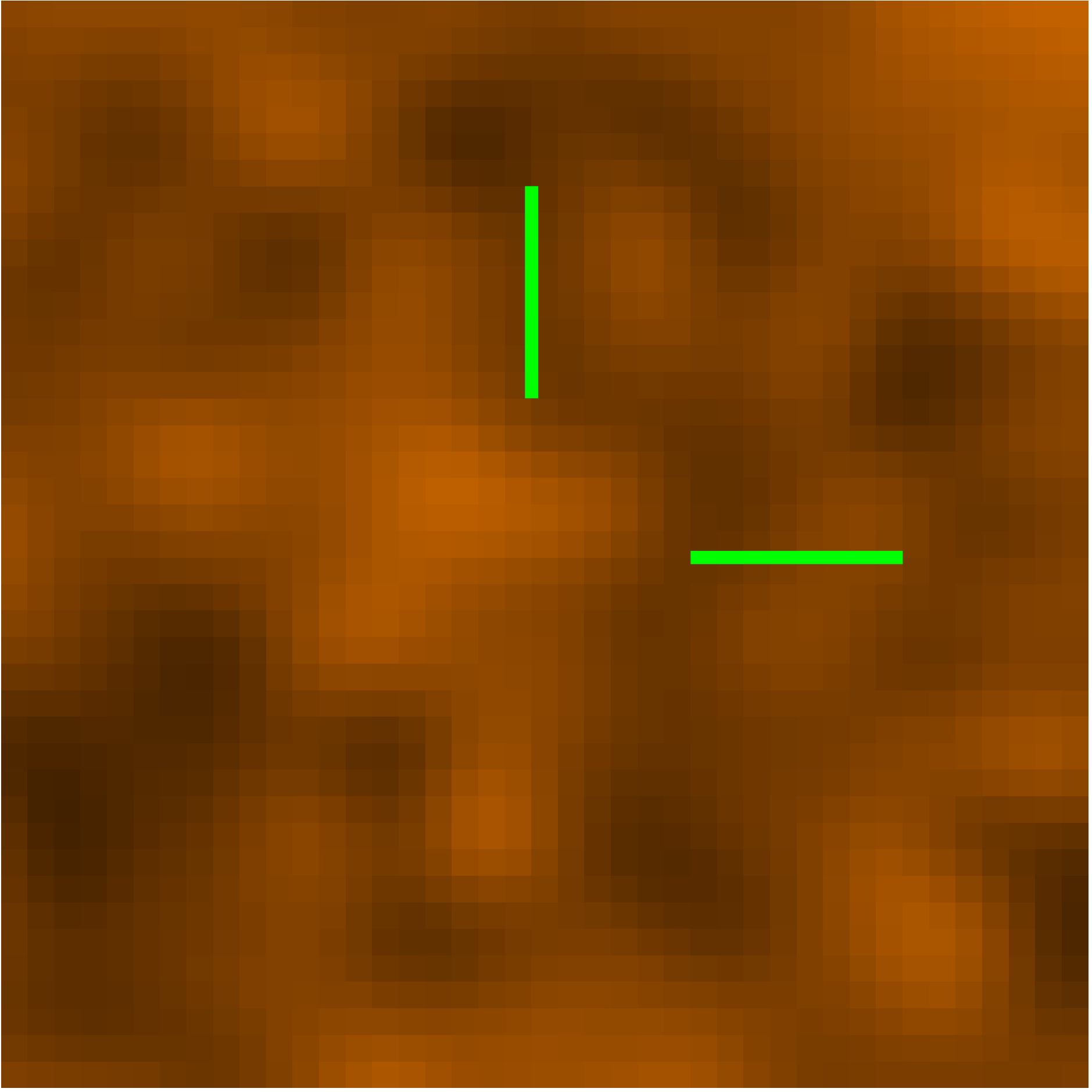}}\\
\subfigure[F160W e1] {\includegraphics[width=0.15\textwidth]{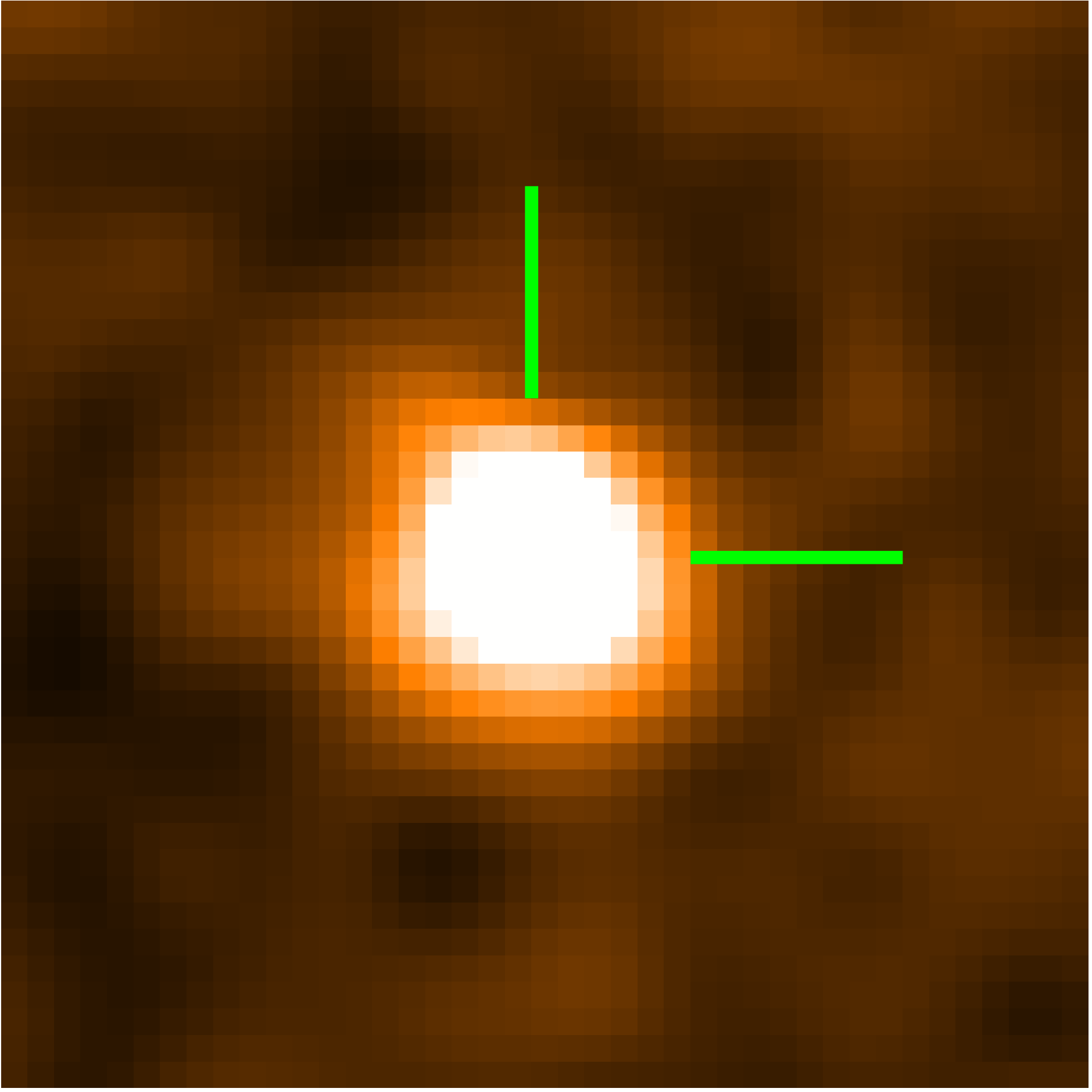}}
\subfigure[F160W e2] {\includegraphics[width=0.15\textwidth]{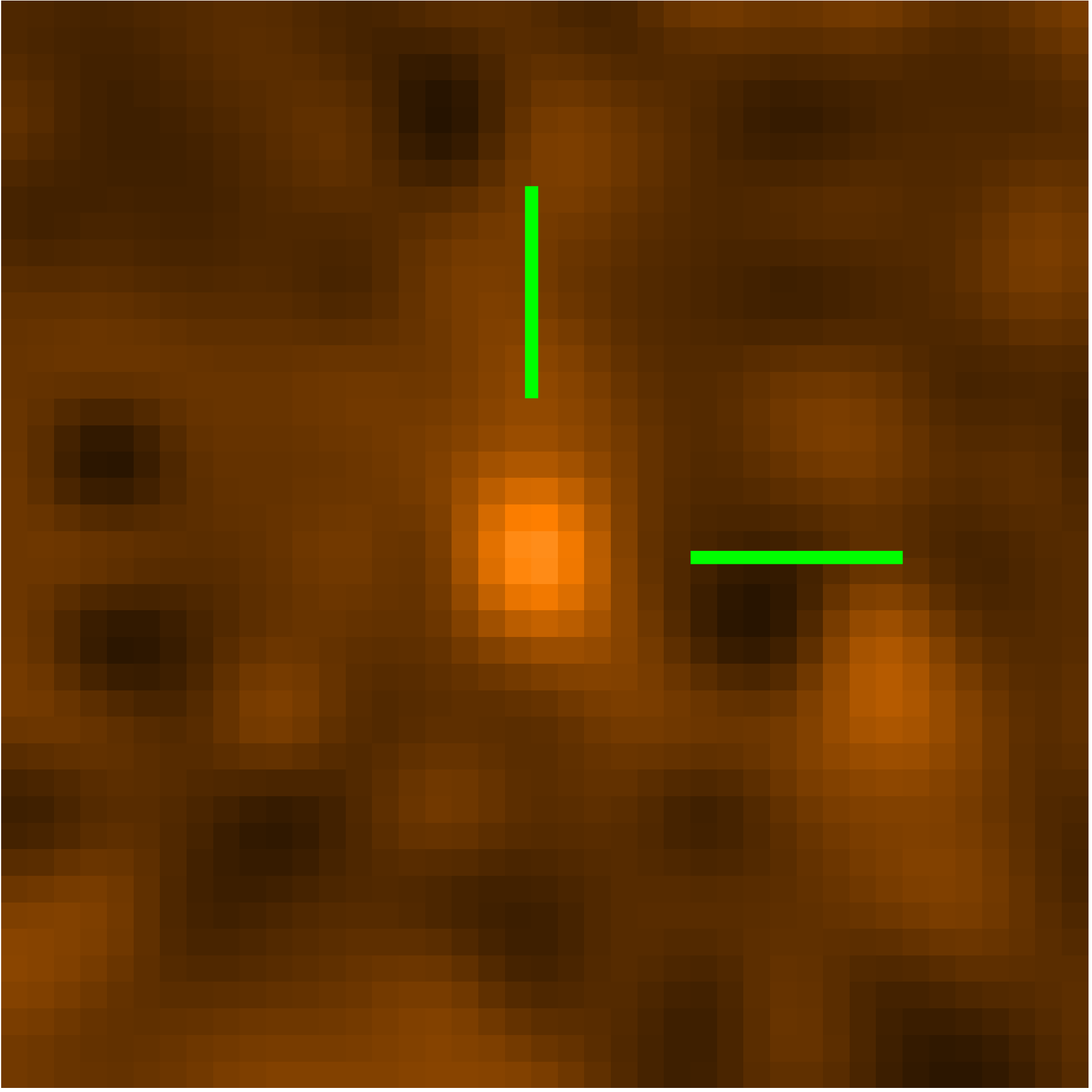}}
\subfigure[F160W ref] {\includegraphics[width=0.15\textwidth]{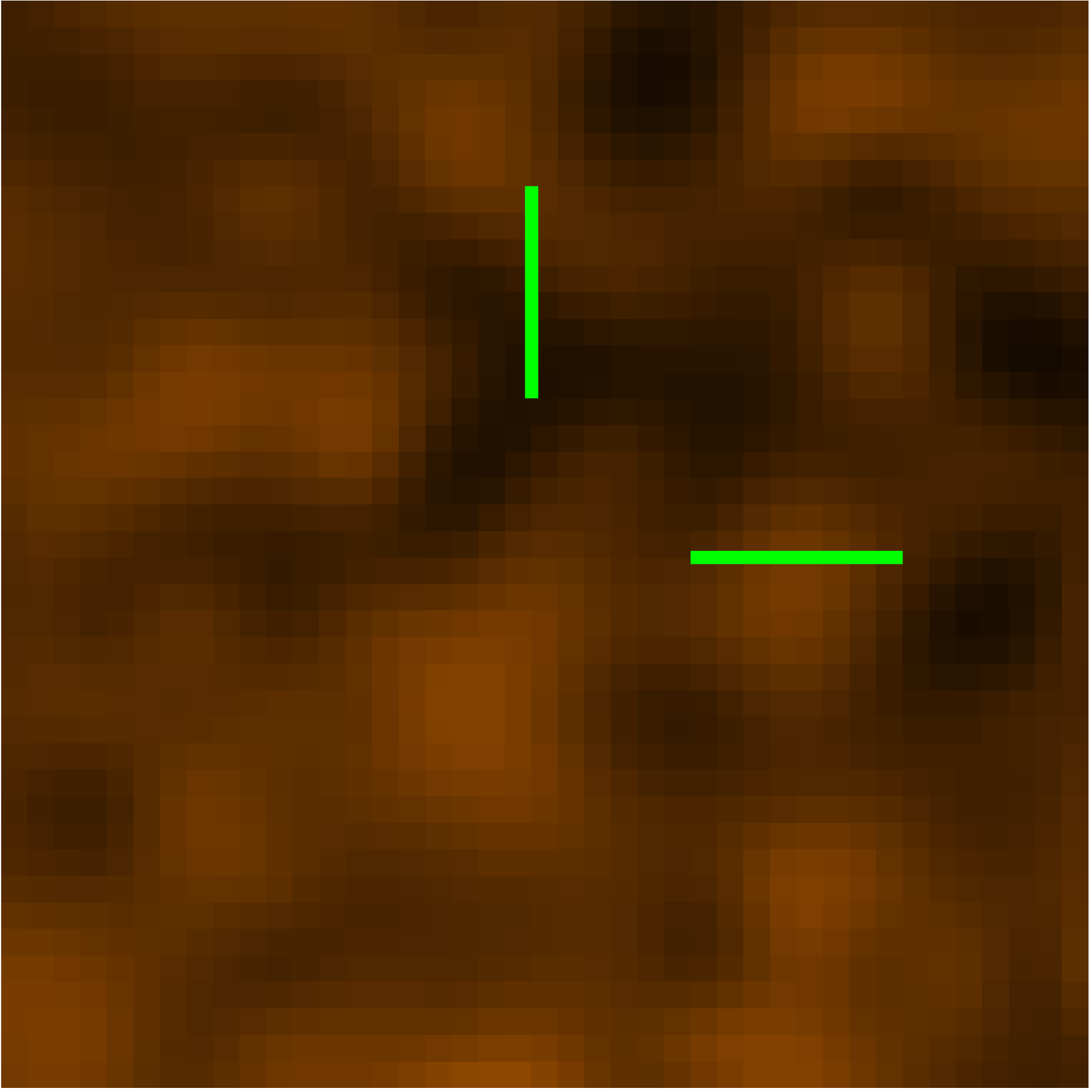}}
\end{center}
\caption{The HST observations of GRB 160821B. The afterglow emission were detected in all bands in the first two epochs of observations,
and faded away in later epochs. In the images of the last epoch, which we take as references, no source is reliably detected near the afterglow position.
}
\label{fig:GRB160821B_field}
\end{figure}

GRB 160821B is located near a bright ($R\sim19.2$ mag) spiral galaxy \citep{Xu2016},
at a redshift of $z=0.16$ \citep{Levan2016}. Deep HST observations found no secure detection of any underlying sources near the afterglow (see Fig.\ref{fig:GRB160821B_field}). We then adopt $z=0.16$ for this burst.  For GRB 160821B, we have adopted a similar analysis method as GRB 150424A, including image subtraction, though there is no underlying source detected. The results are summarized in Tab.2.

\subsection{Identification of the post-jet-break decline behaviors of GRB 150424A and GRB 160821B}\label{subsection:jet}

In the GRB jet model (see Tab.1 of \citet{Zhang2004} for a comprehensive summary), the afterglow lightcurve declines as $t^{-p}$ (or slightly shallower if the
sideways expansion of the ejecta is ignorable), as long as the ejecta has been decelerated to a bulk Lorentz factor $\leq 1/\theta_{\rm j}$,
no matter whether the observer's frequency $\nu_{\rm obs}$ is between $\nu_{\rm m}$ and $\nu_{\rm c}$ (the spectrum is $\propto \nu^{-(p-1)/2}$)
or above both of them (the spectrum is $\propto \nu^{-p/2}$). Therefore, the quick decline $t^{-\alpha}$ at late time, together with a reliable spectral index which is close to $(\alpha-1)/2$ or $\alpha/2$ for $\alpha \gtrsim 2$, are widely taken as the strong evidence for the post-jet-break decline behavior.

\begin{figure}[h]
\centering
\includegraphics[width=0.8\textwidth]{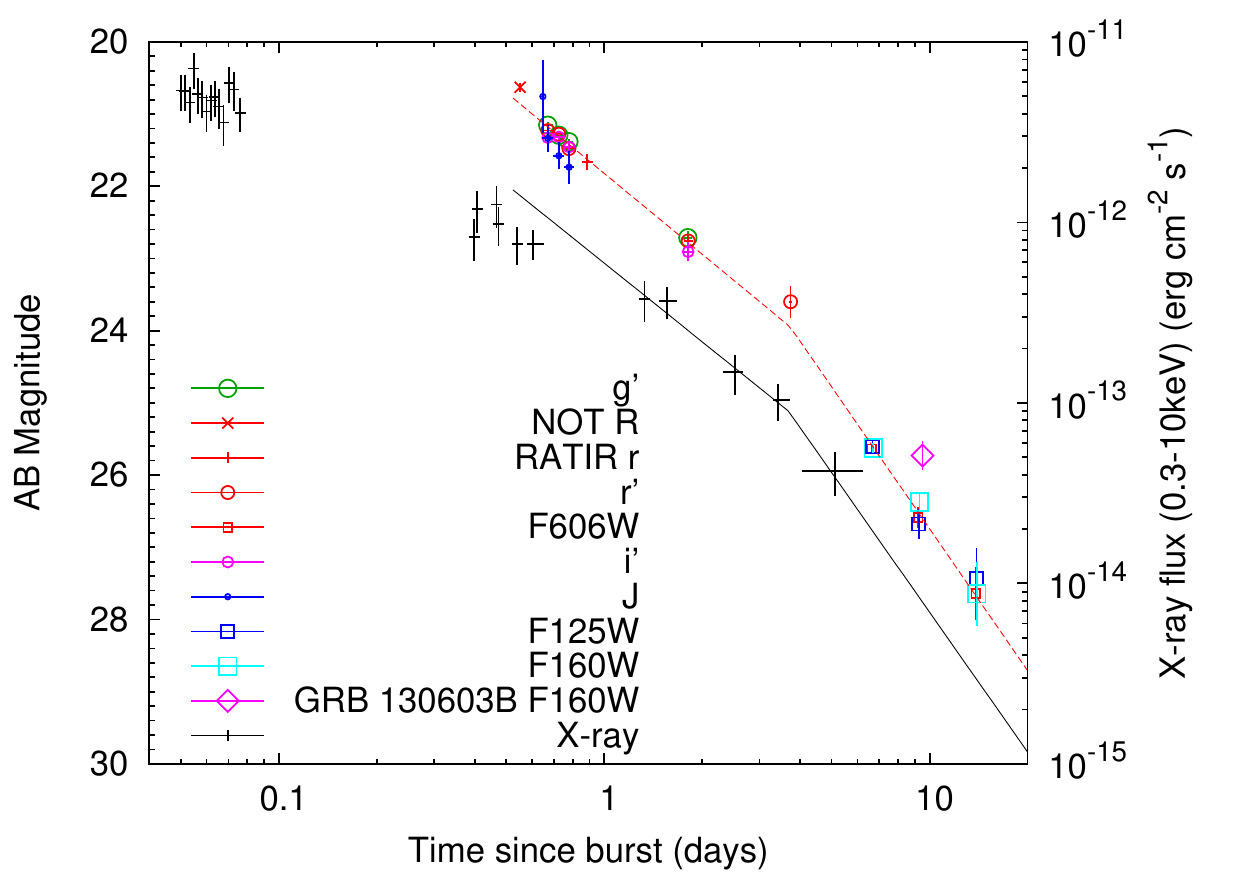}
\caption{The multi-band afterglow lightcurves of GRB 150424A.  The $g', r', i', J$ band data are taken from \citet{Knust2017}, the HST data are analyzed in this work, all bands have been converted to $i'$ band assuming a $\beta=0.75$ power law spectrum {(note that the {\it Swift} UVOT and GROND spectra are similar to the HST spectrum, as shown in Figure 5)}. Dashed line is a broken power-law fit to the lightcurve.
Our fit does not include the NOT $R$ band \citep{Malesani2015} and RATIR $r$ band \citep{Butler2015} data from GCN.
The {\it Swift} XRT lightcurve is provided by the UK {\it Swift} Science Data Centre \citep{Evans2009}, the black line is extrapolated from $i'$ band also with a $\propto \nu^{-0.75}$  spectrum, it is also in agreement with the data.
}
\label{fig:GRB150424A_LC}
\end{figure}

\begin{figure}[h]
\centering
\includegraphics[width=0.8\textwidth]{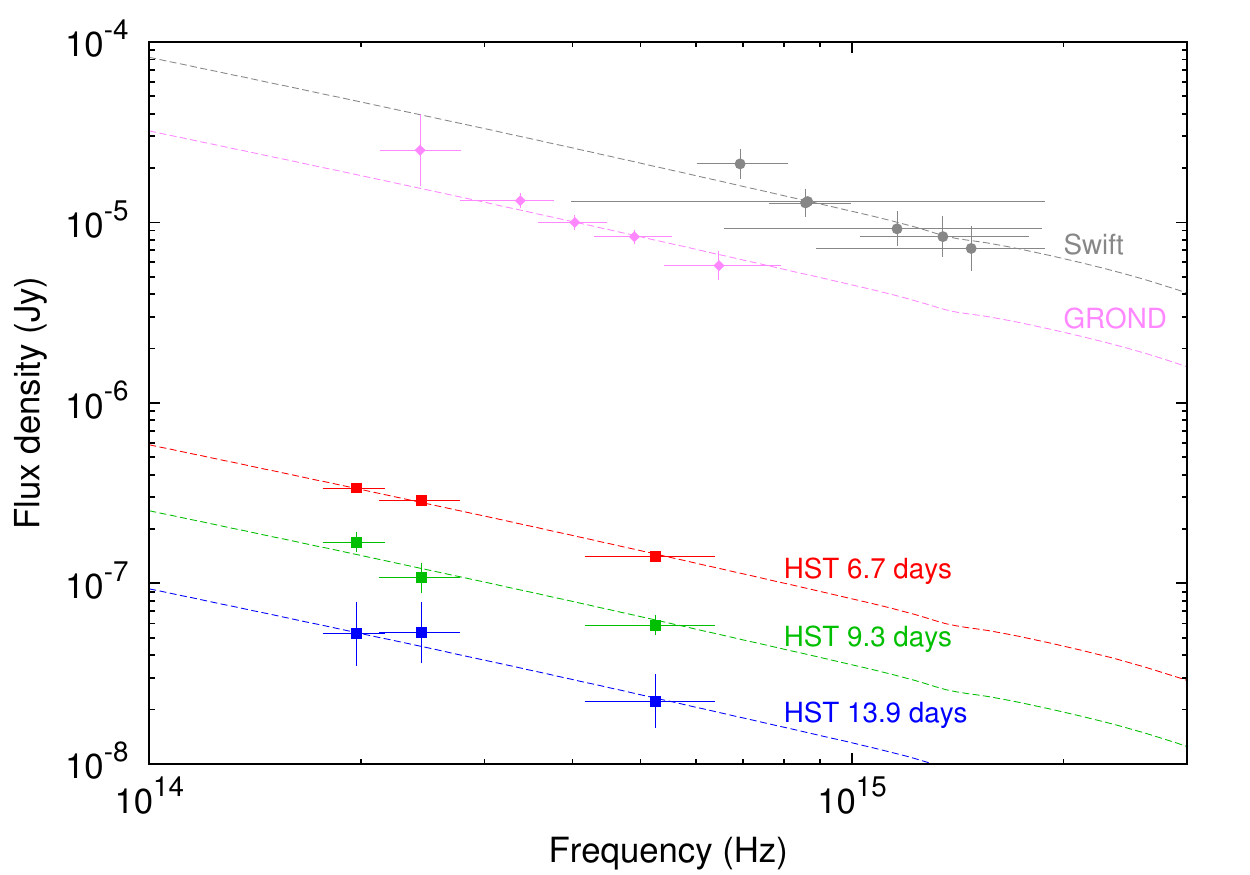}
\caption{The optical SED of the afterglow of GRB 150424A. The HST data are analyzed in this work, they are fitted by a power-law spectrum with extinction of the Galaxy $E(B-V)=0.06$ \citep{Schlafly2011}.  Extinction of the host galaxy is not considered here because there is no significant evidence for such an effect.
The {\it Swift} UVOT data ($uvw2$, $uvm2$, $uvw1$, $white$, $u$ and $b$ bands) are adopted from \citet{Marshall2015},
the GROND data {($g'$, $r'$, $i'$, $z'$ and $J$ bands)} are taken from \citet{Kann2015},
and their power-law spectra are similar to the HST's. {Note that the UVOT data were not measured simultaneously. However, the afterglow emission changed very slowly at early times, as shown in Knust et al. (2017). Moreover, the UVOT data reported in \citet{Marshall2015} are combinations of images during similar time intervals. It is thus reasonable to use these UVOT data to construct a SED.}
}
\label{fig:GRB150424A_SED}
\end{figure}

For GRB 150424A, the F606W, F110W and F160W data are consistent with the post jet break afterglow model for $p\sim2.5$, for which the spectrum should be $\propto \nu^{-0.75}$ (for $\nu_{\rm m}<\nu_{\rm obs}<\nu_{\rm c}$) and the decay should be $\propto t^{-2.5}$, see Fig.\ref{fig:GRB150424A_LC} and Fig.\ref{fig:GRB150424A_SED}. Converting all bands to $i'$ band with a spectrum of $\nu^{-0.75}$, then fitting\footnote{Here and after, we have fitted the data by the minimal $\chi^2$ method, using the MINUIT package (http://lcgapp.cern.ch/project/cls/work-packages/mathlibs/minuit/index.html), and the errors are derived by the MINUIT. The $\chi^2$ is calculated as $\sum (\frac{measured~value - predicted~value}{measured~error})^2$, where the $predicted~value$ is the average flux/magnitude of the predicted light curve within each time bin.} the GROND and HST data with a broken power-law decay, we find a break at $t=3.73\pm0.16$ days, and the decay indices are $1.49\pm0.04$ and $2.61\pm0.12$, respectively. This fit yields a total $\chi^{2}/{\rm d.o.f}=27.8/22$, indicating a reasonable fit for both spectral and temporal behavior of the afterglow. Note that the pre-break decline behavior is also well consistent with the theoretical model (for $p=2.5$, it is $(3p-2)/4\sim1.4$). The {\it Swift} XRT PC mode spectrum, although took earlier than HST observations, gives a photon index of $2.00^{+0.19}_{-0.18}$ \citep{Melandri2015}, which is  consistent with $p\sim 2.5$ supposing the X-ray band is above both $\nu_{\rm m}$ and $\nu_{\rm c}$ (i.e., the spectrum should be $\propto \nu^{-p/2}$). The data strongly prefer a broken power-law model to a single power-law model, by which the fit gives a $\chi^{2}/{\rm d.o.f}=97.5/24$. \citet{Knust2017} identified an optical plateau lasting about $\sim~8$ hours and attributed such a shallow decline to the energy injection from a central millisecond magnetar. This is not at odds with our jet break interpretation because at late times (i.e., when the magnetar spined down and the energy injection rate drops with time as $t^{-2}$) the energy injection was not efficient any longer and the forward shock emission should be rather similar to the normal scenario (i.e., without energy injection; as shown for example in \citet{ZhangB2004}).

\begin{figure}[h]
\centering
\includegraphics[width=0.8\textwidth]{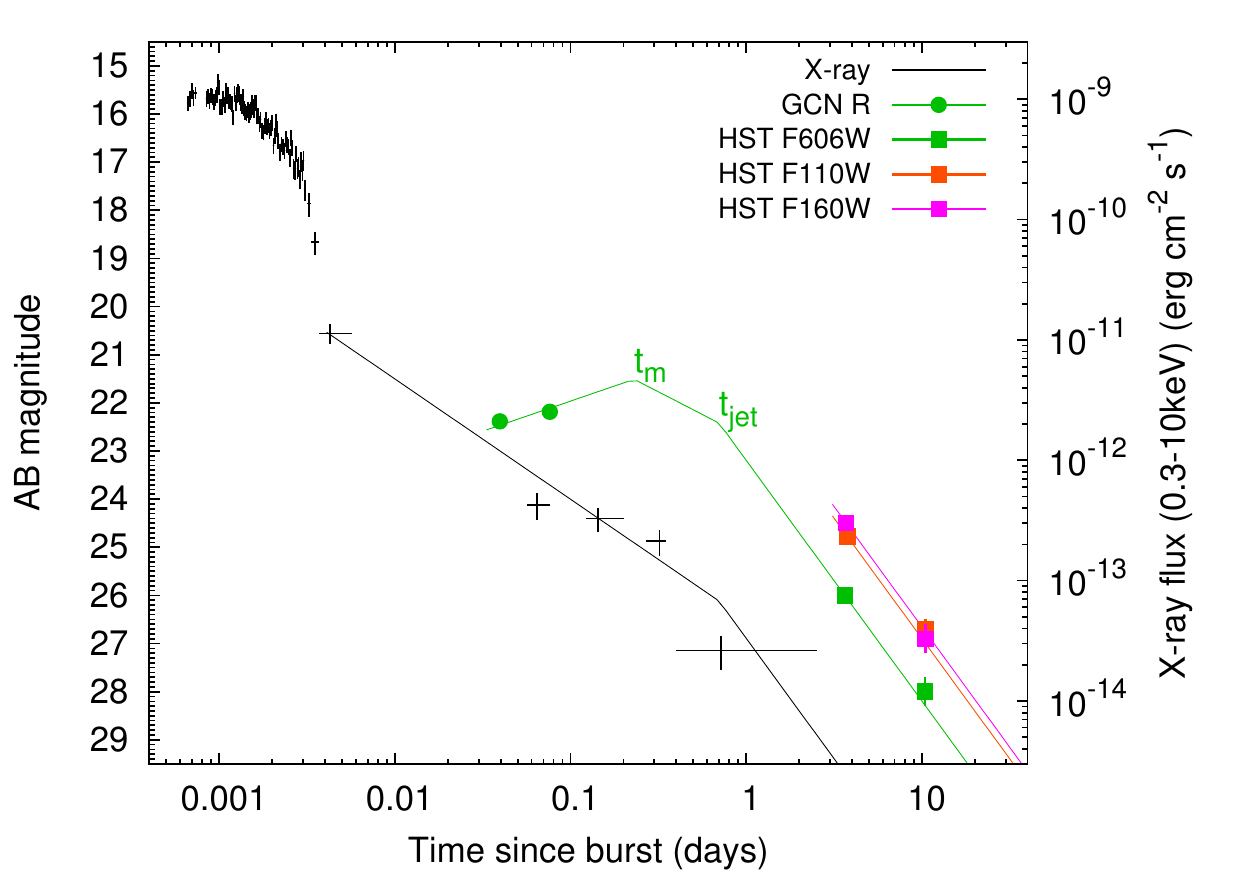}
\caption{The multi-band afterglow lightcurves of GRB 160821B. HST data can be fitted by a power-law decay with index $1.83\pm0.13$
($\chi^{2}/{\rm d.o.f}=1.65/(6-4)$).
The $R$ band data are from NOT \citep{Xu2016} and GTC \citep{Jeong2016},
the {\it Swift} XRT lightcurve is provided by the UK {\it Swift} Science Data Centre \citep{Evans2009}. Though rather sparse,
they can be interpreted by a simple analytic afterglow model (green and black lines). The forward shock emission was in the slow cooling phase and the spectral index of the accelerated electrons ($p$) is about 2. For $t<t_{\rm m}$, the R-band is below $\nu_{\rm m}$ and the flux increases with the time as $t^{1/2}$. At $t=t_{\rm m}$,  $\nu_{\rm m}$ crossed the R-band, then the flux drops with the times as $t^{-3(p-1)/4}$. At $t>t_{\rm jet}$, the flux declines as $t^{-p}$ (the spectra at $t\geq 3.6$ days however suggest that at late times the optical bands are below both $\nu_{\rm m}$ and $\nu_{\rm c}$). As for the X-ray emission, the observer's frequency is above both $\nu_{\rm m}$ and $\nu_{\rm c}$, and the decline is $\propto t^{-(3p-2)/4}$ for $t<t_{\rm jet}$ and $\propto t^{-p}$ for $t>t_{\rm jet}$.
}
\label{fig:GRB160821B_LC}
\end{figure}

For GRB 160821B, the publicly-available data are rather limited. The HST took two epochs of observations at 3.6 days and 10.4 days after the burst, when the afterglow was still visible. Fitting $F606W$, $F110W$ and $F160W$ bands with a same power law decay yields a flux decline $\propto t^{-1.83\pm0.13}$ ($\chi^{2}/{\rm d.o.f}=1.65/2$), see Fig.\ref{fig:GRB160821B_LC}.
 Fitting the $F606W$, $F110W$ and $F160W$ band SED with a power law spectrum, we find the power law indices are $\alpha=1.35\pm0.07$ ($\chi^2/{\rm d.o.f}=13/1$) at $t=3.6$ days and $\alpha=1.12\pm0.39$ ($\chi^2/{\rm d.o.f}=3.36/1$) at $t=10.4$ days, respectively. Alternatively, if we fit the data with a thermal spectrum, the resulting temperature is of $4488\pm81[(1+z)/1.16]$ K ($\chi^2/{\rm d.o.f}=15/1$) at $t=3.6$ days and $=4750\pm473 [(1+z)/1.16]$ K ($\chi^2/{\rm d.o.f}=0.30/1$) at $t=10.4$ days, respectively. Though a thermal spectrum can well reproduce the SED at $t=10.4$ days, the temperature is much higher than $\sim 2500$ K, the value inferred for the macronova signals of GRB 060614 and GW170817/GRB 170817A in a similar epoch \citep[see][respectively]{Jin2015,Drout2017}. Instead, the overall decline behavior as well as the power-law spectrum at $t=10.4$ days are largely consistent with the jet model for $p\sim 2$ and $\nu_{\rm obs}>\max\{\nu_{\rm m},\nu_{\rm c}\}$ (the spectrum is $\propto \nu^{-p/2}$). Moreover, the {\it Swift} XRT spectrum measured in the time interval of $t\sim 4873-47180$ s has a power-law photon index $2.0^{+0.7}_{-0.6}$ (http://www.swift.ac.uk/xrt$_{-}$spectra/00709357/; Evans et al. 2009), which is consistent with $p\sim 2$, too. Supposing the sideways expansion of the ejecta is unimportant, at early times the flux declines as $t^{(2-3p)/4}$ while in the post jet-break phase the decline is steepened by a factor of $t^{-3/4}$, as shown in \citet{Zhang2004}.
 Therefore, we suggest that the HST data of GRB 160821B were dominated by a power-law afterglow component though at $t=3.6$ days an underlying weak macronova component is possible (please see Sec.\ref{subsection:macronova} for further discussion).
 The X-ray afterglow emission in the time interval of $\sim 0.05-2$ day after the burst also strongly indicate the presence of a jet break at $t\geq 0.3$ day. The analytical approach yields a $t_{\rm jet}\sim 0.7$ day (see Fig.\ref{fig:GRB160821B_LC}), which is comparable to though a bit larger than the X-ray data based estimate by \citet{Lv2017}.

With the break time, the jet half-opening angle can be estimated by \citep{Sari1999,Frail2001}
\begin{equation}
        \theta_{\rm j} \approx 0.076
              \Bigl({t_{\rm jet} \over 1\, {\rm day}}\Bigr)^{3/8}
              \Bigl({1+z \over 2}\Bigr)^{-3/8}
              \Bigl({E_{\rm iso} \over 10^{51}\,{\rm erg}}\Bigr)^{-1/8}
              \Bigl({\eta_\gamma \over 0.2}\Bigr)^{1/8}
              \Bigl({n \over 0.01\,{\rm cm^{-3}} }\Bigr)^{1/8}.
\label{eq:theta-j}
\end{equation}
For GRB 150424A, Konus-Wind recorded a total gamma-ray (20 keV$-$10 MeV) fluence of $1.81\pm0.11\times10^{-5}~{\rm erg}~{\rm cm}^{-2}$ \citep{Golenetskii2015},
corresponding to an isotropic gamma-ray emission energy of $E_{\rm iso}\sim4.3\times10^{51}$ or $\sim1.0\times10^{53}$ erg at $z=0.3$ or $1.0$, respectively,
using cosmological parameters of Planck results \citep{Planck2014}.
The Fermi gamma-ray ($10-1000$ keV) fluence is  $1.68\pm0.19\times10^{-6} {\rm erg}~{\rm cm}^{-2}$ for GRB 160821B \citep{Stanbro2016},
so we have $E_{\rm iso}\sim 10^{50}~{\rm erg}$ for $z=0.16$. Following \citet{Frail2001} we take the radiation efficiency $\eta_\gamma=0.2$. For the ISM number density $n=0.01~{\rm cm^{-3}}$ (see Fong et al. 2015 for the evidence of the low number density of the ISM surrounding SGRBs), we get $\theta_{\rm j}\approx 0.12$ for GRB 150424A (if we take $z=1.0$ then $\theta_{\rm j}\sim 0.07$) and $\theta_{\rm j}\approx 0.1$ for GRB 160821B.

A reliable jet half-opening angle has only been rarely inferred for SGRBs/lsGRBs (Please see Tab.3 for a summary).
Interestingly, all low-redshift ($z\leq 0.4$) events (except GRB 050502B, of which the optical afterglow emission was never detected) with deep HST follow-up observations
are in this sample. These events include GRB 050709, GRB 060614, GRB 130603B, GRB 150424A and GRB 160821B (Note that for GRB 050709, no jet break was directly measured. But the presence of a macronova signal strongly favors a jet break at $t\leq 1.4$ days after the burst, see Jin et al. 2016). Such an observational fact likely points towards the presence of jet break in most events and their non-detection may simply due to the lack of deep follow-up observations. The other interesting feature is the ``narrow" distribution of these estimated $\theta_{\rm j}$ that peaks at $\sim 0.1$ rad, indicating that many more SGRBs/lsGRBs were off-beam (or off-axis).

\begin{table}
\begin{center}
\label{tab:SGRBjet}
\title{}Table 3. Short/lsGRBs with a jet break (or upper limit). \\
\begin{tabular}{cccccc}
\hline
\hline
GRB & z	& $E_{{\rm iso}}$	& $t_{\rm jet}$	& $\theta_{\rm j}$	& References \\
 	& 	&	($10^{51}$erg)	& (days)	& (rad)	& \\
\hline
050709	& 0.16	& 0.07	& $< 1.4 $	& $ <0.14 $ & (1) \\
051221A	& 0.546	& $2.4$	& 5 & 0.09 & (2,3)              \\
060614	& 0.125	& 2.5	& 1.4 & 0.08-0.09	& (4,5)        \\
061201  & 0.111 &  0.14  &   0.03  & 0.02-0.03  &  (6)\\
090426A	& 2.609	& $4.2$	& 0.4 & 0.08-0.12 & (7)       \\
111020A	& --	& $0.21/1.9^{a}$	& 2 & 0.05-0.14	& (8) \\
130603B	& 0.356	& 2.1	& 0.47	& 0.07-0.14 & (9,10) \\
140903A	& 0.351	& $0.06$	& 1.2	& 0.05-0.1	& (11, 12) \\
150424A	& 0.30	& 4.3	& 3.7	& 0.12	&  this work   \\
160821B	& 0.16	& 0.21	& 0.7	& 0.10	&  this work,(13)   \\
\hline
\end{tabular}
\end{center}
{a. The isotropic-equivalent prompt emission energy was roughly estimated in eq.(8) by assuming different redshifts and physical parameters.
References: (1) \citet{Jin2016}; (2) \citet{Soderberg2006};(3)\citet{Burrows2006}; (4) \citet{Xu2009};(5) \citet{Mangano2007}; (6) \citet{Stratta2006}; (7) \citet{NicuesaGuelbenzu2011}; (8) \citet{Fong2012}; (9) \citet{Fong2014}; (10) \citet{Fan2013};
(11) \citet{Troja2016a}; (12) \citet{ZhangS2017}; (13) \citet{Lv2017}.}
\end{table}

\subsection{Any macronova signal in GRB 150424A and/or GRB 160821B?}\label{subsection:macronova}

For GRB 150424A, both the temporal and spectral behaviors are well consistent with the afterglow model and there is no macronova signal,
see Fig.\ref{fig:GRB150424A_LC} and Fig.\ref{fig:GRB150424A_SED}.

\begin{figure}[h]
\centering
\includegraphics[width=0.5\textwidth]{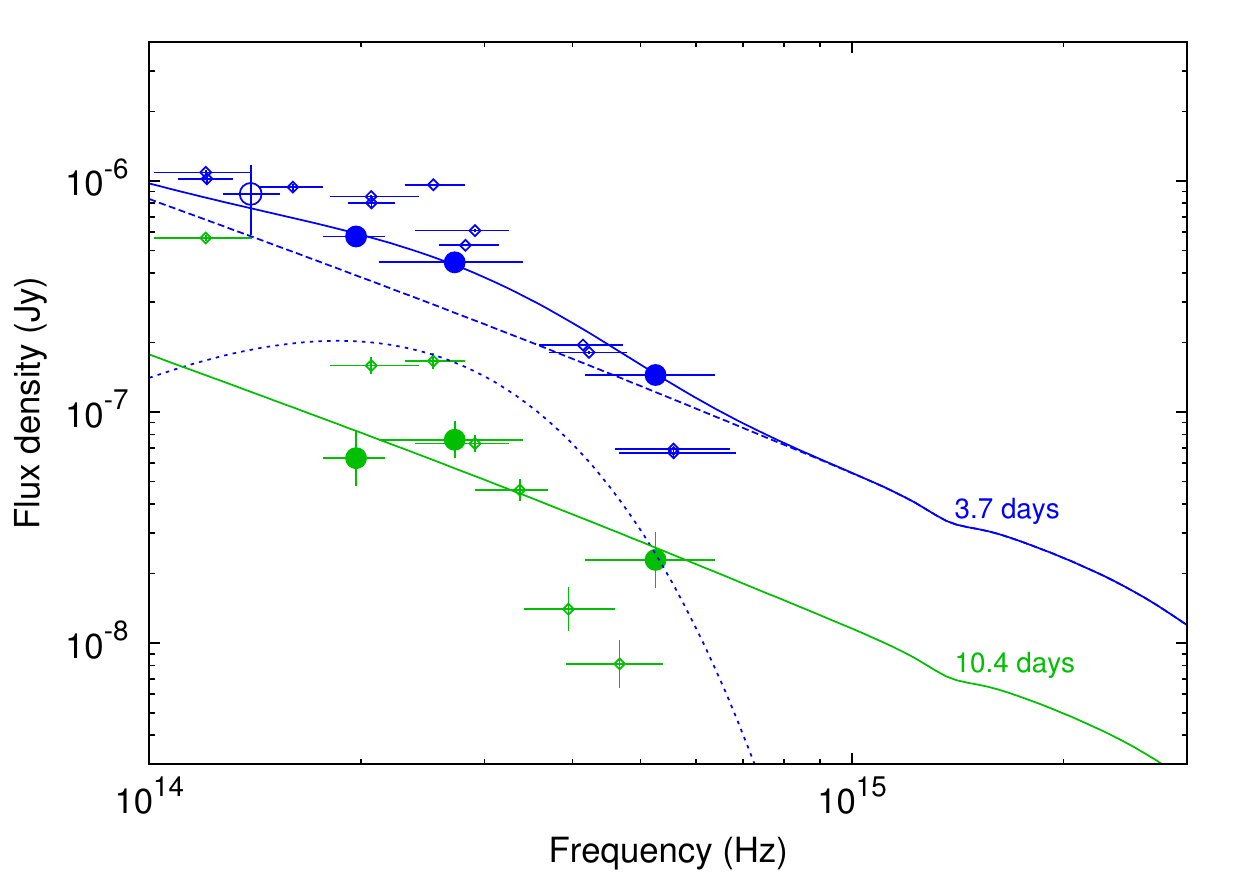}
\caption{The SED of GRB 160821B. Big solid circles represent HST  $F606W$, $F110W$ and $F160W$ band data measured at 3.6 days and 10.4 days after the burst, and the single big empty circle is the Keck $K_{\rm s}$ band data measured at 4.3 days \citep{Kasliwal2017a}. Since the spectrum at $t\sim 3.6$ days can not be reasonably fitted by a single power, here we fit the data by a power-law component (dashed lines) plus a weak thermal (dotted lines) component. Due to the rarity of the data, the power law index and the temperature are fixed to 1.1 and 3100 K. The Galactic extinction of $E(B-V)=0.04$mag \citep{Schlafly2011} has been taken into account in the fit, while the extinction of the host galaxy is not considered because of the lack of relevant data. The SED has also been compared with AT2017gfo, the macronova associated with the gravitation wave event GW 170817, which has been shifted from $z=0.0095$ to $z=0.16$. The data of AT2017gfo (in diamond) are from VLT \citep{Pian2017}, Gemini \citep{Kasliwal2017b,Troja2017} and VISTA \citep{Tanvir2017}, and corrected for the Galactic extinction of $E(B-V)=0.11$\citep{Schlafly2011}. Although the afterglow flux of GRB 160821B in $F110W$ and $F160W$ bands are comparable with AT2017gfo (shifted to $z=0.16$), their SEDs are however quite different (i.e., the latter is much softer). Thus we suggest that the afterglow of GRB 160821B is dominated by a power law spectrum component, but a weak underlying macronova component with a flux of $\sim 1/5$ that of AT2017gfo is possible.} \label{fig:GRB160821B_SED}
\end{figure}

As for GRB 160821B, the situation is less clear. The HST data at $t\sim 3.6$ days are hard to be interpreted as either a power-law or a thermal spectrum (see Sec.\ref{subsection:jet}). Interestingly, they can be interpreted as the superposition of a power-law afterglow component (with a spectrum $f_{\nu} \propto \nu^{-1.1}$) and a thermal-like component with a temperature of $\sim 3100$ K (due to the sparse data points, this temperature is fixed to be that of AT2017gfo in the same epoch\footnote{One can also assume that at $t=10.4$ days, there is also a thermal component at a temperature of $\sim 2500$ K. A $f_{\nu} \propto \nu^{-1.1}$ power-law spectrum plus such a thermal component fit finds out that the thermal component has a flux lower than the data points by a factor of $\sim 10$. Indeed, as revealed in AT2017gfo, the late time optical/infrared macronova emission \citep[e.g.][]{Pian2017,Kasliwal2017b,Tanvir2017} drops with time quicker than $t^{-3}$.}) and the fit yields $\chi^{2}/{\rm d.o.f}=1.36/1$, as shown in Fig.\ref{fig:GRB160821B_SED}. Such a result is likely in support of \citet{Troja2016b}'s speculation that there might be a weak macronova signal in the afterglow of GRB 160821B. Though intriguing, the current limited publicly-available HST data alone are insufficient to draw a final conclusion (T. Piran 2017, private communication).

\subsection{The neutron star merger rate in the local ($z\leq 0.2$) universe}

The SGRB-data based neutron star merger rate was widely estimated in the literature \citep[e.g.,][]{Guetta2005,Nakar2007,Coward2012,Fong2013,Petrillo2013,Fong2015}. However, all these previous approaches were { mainly} based on jet opening angles measured at relatively high redshifts {(though GRB 050709, GRB 060614 and GRB 061201 were included ``individually" in some estimates)}. The reasonable estimate of $\theta_{\rm j}$ for GRB 050709, GRB 060614, GRB 061201 and GRB 160821B, four GRBs at redshifts of $\leq 0.2$, provides the first opportunity to directly though conservatively estimate the neutron star merger rate in the local universe that can be directly tested by (compared to) the ongoing advanced LIGO/Virgo observations.

GRB 050709 was detected by HETE-II with a field of view ${\rm F.o.V}\approx 3\,{\rm sr}$  \citep{Villasenor2005}, while GRB 060614, GRB 061201 and GRB 160821B were recorded by {\it Swift} satellite with a ${\rm F.o.V}\approx 1.4-2.4\,{\rm sr}$  \citep{Gehrels2006,Siegel2016}. Note that HETE-II has a much lower detection rate of SGRBs in comparison with {\it Swift} due to the relatively small effective area of the onboard detector. Moreover, no GRBs were reported by HETE-II any longer since March 2006 (http://space.mit.edu/HETE/Bursts/). The joint analysis of HETE II and {\it Swift} SGRBs are thus very challenging. In this work for simplicity we exclude GRB 050709 in the
following investigation.

Now we evaluate the ``apparent" (i.e., without the jet half-opening angle correction) rate, ${\cal R}_{\rm nsm,app}$, for local SGRBs. {The apparent rate is related to the observational number of events as well as the sensitivity and F.o.V of the instrument. For a given redshift, the sensitivity plays a role in determining the weakest SGRB that can be detected by the instrument while the F.o.V reflects the search portion of the full sky. The BAT onboard {\it Swift} is a coded mask telescope, and its F.o.V and sensitivity change as a function of the partial coding fraction, which is related to the burst's incident angle \citep{2005SSRv..120..143B}; meanwhile, BAT has a complex trigger algorithm, making detailed analysis on the intrinsic rate through {\it Swift} data very complicated \citep{2014ApJ...783...24L}. {In this work we take into account these effects as the following: \citet{2016ApJ...829....7L} showed that the sensitivity of BAT decreases with the burst's duration (which roughly reflects the exposure time) with ${\rm flux}_{\rm limit} \propto 1/\sqrt{T_{\rm 90}}$; they also showed that the sensitivity to 1s flux is $\sim 3 \times 10^{-8} \rm{erg~cm^{-2}~s^{-1}}$ for fully-coded region. If a burst's incident angle is large and  the detector's plane is partially coded, the effect of partial coding fraction $p_{\rm f}$ can be expressed with the ``effective on-axis exposure time" $T_{\rm eff} = p_{\rm f}T_{\rm 90}$ \citep{2013ApJS..207...19B}. With these relations, we collect the 1s peak fluxes of the three bursts from the {\it Swift}/BAT Gamma-Ray Burst Catalog, and calculate the corresponding smallest partial coding at different distances by
 \begin{equation}
p_{\rm f}= \left [ \frac{3\times 10^{-8} \rm{erg/cm^{2}/s}}{\left (\frac{D_{\rm L0}}{D_{\rm L}}  \right )^{2}\left ( \frac{1+z}{1+z_{0}} \right )f_{0}} \right ]^{2}/T_{90}
 \end{equation}
where $z_{\rm 0}$, $D_{\rm L0}$ and $f_{\rm 0}$ are the observed redshift, luminosity distance, and 1s peak flux of a given burst. The yielded ${p}_{\rm f}$
is then used to calculate the corresponding BAT F.o.V, and their relation are inferred from
\citet{2005SSRv..120..143B} (we fit their simulated curve adjusted for off-axis projection effects with polynome and obtain a analytical function ${\rm F.o.V}(p_{\rm f})$). The F.o.V is then adopted to calculate the space-time volume of the search for a given burst:
 \begin{equation}
 \left\langle {VT} \right\rangle  = 0.9T\int_0^{0.2} {\frac{F.o.V\left ( {p}_{\rm f} \right )}{4\pi }\frac {1}{1 + z}\frac{d{V_c}\left( z \right)}{dz}dz}
 \end{equation}
where the factor 0.9 represents the fraction of the time that BAT spends on searching for GRBs. Assuming a negligible evolution of rate in the local universe, and the observed number of event in a given space-time volume should follow a Poission distribution, we use the Bayesian inference to derive the posterior distribution of ${\cal R}_{\rm nsm,app}$ for each burst by \citep[see][and the references therein]{WangYZ2017}
 \begin{equation}
P({\cal R}_{\rm nsm,app}) \propto P_{\rm Poission}(1|{\Lambda}) \times {P}'({\cal R}_{\rm nsm,app}),
\end{equation}
where $P_{\rm Poission}(1|{\Lambda})$ is the likelihood of observing one event from a Poission distribution with a mean number $\Lambda = {\cal R}_{\rm nsm,app}{\left\langle {VT} \right\rangle}$, and $P'({\cal R}_{\rm nsm,app})$ is the prior (for which we choose a uniform distribution). The inferred rates from GRB 060614, GRB 061201 and GRB 160821B are $ 0.25^{+0.37}_{-0.18},0.26^{+0.40}_{-0.19}$ and $0.30^{+0.45}_{-0.22}~{\rm Gpc^{-3}yr^{-1}}$, respectively. In this work the errors stand for the 68\% credible intervals unless specifically noted. These rates are similar, indicating that these three bursts are bright and hence the influence from the change of F.o.V is not significant.}


To get the local neutron star merger rate, the geometry correction (including the uncertainties of $\theta_{\rm j}$) should be addressed. We assume that the probability distribution of the true values of $\theta_{\rm j}$ of GRB 060614 and GRB 061201 follow uniform distribution in the intervals reported in Table 3. For GRB 160821B, 
we assume its $\theta_{\rm j}$  distributes uniformly within $0.08-0.18~{\rm rad}$ since $t_{\rm jet}$ should be within $\sim 0.3-3$ days after the burst (see the X-ray and optical data in Fig.6). The posterior distribution of local neutron star merger rate derived from each burst, ${\cal R}_{\rm nsm}$, is then calculated by
\begin{equation}
p\left ( {\cal R}_{\rm nsm} \right ) \propto \int_{\theta_{\rm j,min}}^{\theta_{\rm j,max}} p\left (1 | {\cal R}_{\rm nsm},\theta_{\rm j}\right ){p}'\left ( {\cal R}_{\rm nsm} \right ){p}'(\theta_{\rm j})d\theta_{\rm j}
\end{equation}
in which the priors for $\theta_{\rm j}$ follow the distributions discussed above, and the prior for ${\cal R}_{\rm nsm}$ is assume to be uniform. The likelihood term can be written as
\begin{equation}
p\left (1 | {\cal R}_{\rm nsm},\theta_{\rm j}\right ) = p_{\rm poisson}\left ( 1 | {\cal R}_{\rm nsm}\frac{{\theta_{\rm j}}^2}{2} \left \langle VT \right \rangle \right  ).
\end{equation}
The local neutron star merger rates inferred from GRB 060614, GRB 061201 and GRB 160821B are $68^{+103}_{-50},832^{+1407}_{-625}$ and $33^{+84}_{-27}~{\rm Gpc^{-3}yr^{-1}}$ respectively. The total rate, dominated by GRB 061201, is obtained by convoluting the three distributions together, with which we have
\begin{equation}
{\cal R}_{\rm nsm}=1109^{+1432}_{-657}~{\rm Gpc^{-3}yr^{-1}}.
\label{eq:Rnsm}
\end{equation}
Correspondingly, the total rate with the 90 percent credible interval is $1109^{+2872}_{-840} {\rm Gpc^{-3}yr^{-1}}$.}

Besides the issues examined above, some other factors may further increase the uncertainties to our results \citep[see e.g.,][]{Coward2012}.
(a) The redshift of GRB 061201 is less secure \citep{Stratta2006} than the other events\footnote{Note that for $z=0.111$ the inferred $\theta_{\rm j}\lesssim 2$ deg is already a few times smaller than the others reported in Table 3. At a redshift $\sim 1$, one has $\theta_{\rm j}\lesssim 0.5$ deg \citep{Stratta2006}, which is hard to understand in the short GRB scenario. Such a fact likely disfavors the high redshift hypothesis. That is why in this work we take the $ {\cal R}_{\rm nsm}$ reported in eq.(\ref{eq:Rnsm}) as the fiducial value though the exclusion of GRB 061201 will yield a significantly lower merger rate.}. Since our sample number is small, this will increase the uncertainty of the result. In the following, we will exclude GRB 061201 and do the calculation again for comparison. (b) Only $\lesssim 1/4$ of the SGRBs have measured redshifts \citep{Berger2014}
and some other SGRBs could be nearby, too\footnote{For the local merger-driven GRBs, the situation will change dramatically in the GW era since the gravitational wave data alone can yield reliable luminosity distance and hence the missing $z$ problem will be solved. The intense deep followup observations of the GW/GRB events are also extremely helpful in measuring the jet breaks of the afterglows.}. For example, recently \citet{Siellez2016} claimed that GRB 070923 and 090417A were at $z<0.1$. Moreover, in our analysis two ``local" bursts (GRB 080905A and GRB 150101B) without a reliable jet opening angles are excluded in the rate estimate. It is also probably that only a fraction of neutron star mergers can produce either SGRBs or lsGRBs. Taking into account these factors, the real neutron star merger rate density should be enhanced by a factor of $f_{\rm c}>1$. As a conservative estimate on ${\cal R}_{\rm nsm}$, we do not correct this further.
(c) There are arguments that GRBs shorter than 2 seconds may have collapsar origin, while bursts with duration longer than 2 seconds may still have non-collapsars origin \citep{Bromberg2013}. This is indeed the case for GRB 060614, which is a long-duration burst but the identified macronova component in its late afterglow \citep{Yang2015,Jin2015} revealed its neutron star merger origin. For the other nearby events (GRB 050709, GRB 061201 and GRB 160821B), no supernova emission were detected down to very stringent limits, which strongly disfavored the collapsar origin.
Therefore for our sample this correction is likely irrelevant. (d) Due to its limited energy range, the SGRB detection rate of {\it Swift} is found to be a factor of $R_{\rm b/s}=6.7$ lower than that of BATSE. \citet{Coward2012} amplified the {\it Swift} SGRB rate by such a factor to crudely correct the bias. Though interesting, it is unclear whether this is a reasonable approximation for the local events that are of our interest. Moreover, such a correction likely enhances the merger rate. In summary, despite all these factors {play} roles in the rate estimation, only point (a) seems to be able to significantly change our ``conservative" estimate. If we exclude GRB 061201 from the sample, {the apparent SGRB rate and neutron star merger rate are $0.81^{+0.60}_{-0.39} {\rm Gpc^{-3}yr^{-1}}$ and
\begin{equation}
{\cal R}_{\rm nsm,w/o~061201}=162^{+140}_{-83} {\rm Gpc^{-3}yr^{-1}},
\label{eq:Rnsmw}
\end{equation} respectively.  The resulting neutron merger rate is significantly smaller than that found in eq.(\ref{eq:Rnsm}), suggesting that our results are sensitively dependent on the narrowly beamed burst GRB 061201 and a larger sample is crucial to get a more reliable estimate on ${\cal R}_{\rm nsm}$.}

{Intriguingly, the successful detection of GW170817 in the O2 run of advanced LIGO yields a neutron star merger rate of} \citep{LVC2017}
\[
 {{\cal R}_{\rm nsm,gw}=1540^{+3200}_{-1220}~{\rm Gpc^{-3}~yr^{-1}},}
\]
{\it which is in agreement with the local SGRB-based neutron star merger rate reported in eq.(\ref{eq:Rnsm}).} For eq.(\ref{eq:Rnsmw}), the consistence with ${\cal R}_{\rm nsm,gw}$ is marginal.

The advanced LIGO detectors can detect the gravitational wave radiation from double neutron star mergers within a typical distance $D\sim 220$ Mpc at its designed sensitivity \citep{Abadie2010}. The detection rate of neutron star mergers is thus
\begin{equation}
  {R}_{\rm gw,nsm}={4\pi D^{3}\over 3} {\cal R}_{\rm nsm,gw} \sim 69^{+143}_{-55}~{\rm yr^{-1}}~({D\over 220~{\rm Mpc}})^{3}.
\label{eq:R_GWnsm}
\end{equation}
Notice that the above number assumes a perfect GW detector, a realistic one operates with a duty cycle and
in reality the expected value would decrease by the coincident duty cycle. Nevertheless, the detection prospect is indeed quite promising and we expect that
much more binary neutron star merger events will be detected in the near future.

\section{GRB/GW association: the contribution of off-beam and off-axis events}
\begin{figure}[h]
\centering
\includegraphics[width=0.8\textwidth]{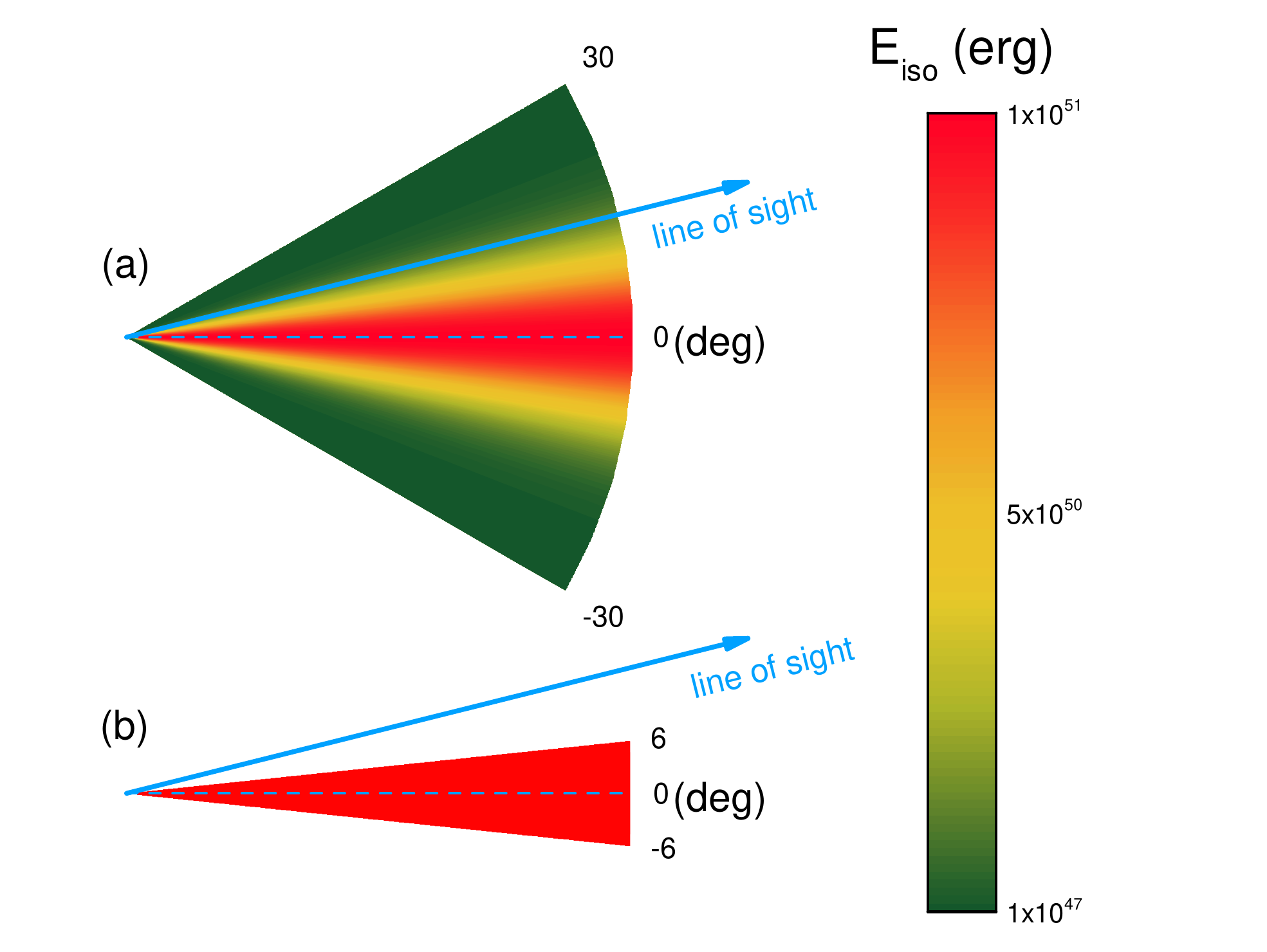}
\caption{(a) Illustration of the off-axis  scenario in the structured jet model. (b)  Illustration of the off-beam  scenario in the uniform jet model.
} \label{fig:cartoon}
\end{figure}

The energy distribution of the GRB ejecta is still unclear. In the standard fireball afterglow model,
a conical jet with a uniform energy distribution within the cone and sharp
energy depletion at the jet edge are assumed (i.e., $\epsilon(\theta)=\epsilon_0$ for $\theta\leq \theta_{\rm j}$ otherwise equals to 0, where $\epsilon(\theta)$ denotes the energy distribution of the ejecta as a function of the polar angle $\theta$). To account for the diverse prompt/afterglow emission of LGRBs, it is further proposed that a non-uniform distribution of energy per solid angle within the jet (i.e., the jets are structured) and the widely-discussed scenarios include (a) the power-law distribution model $\epsilon(\theta)=\epsilon_0 \theta^{-2}$ for $\theta>\theta_{\rm c}$ \citep{Rossi2002,ZhangB2002,Dai2001} otherwise $\epsilon(\theta)=\epsilon_0$; (b) the Gaussian-type jet $\epsilon(\theta)=\epsilon_0 \exp{(-\theta^2/2\theta_{\rm c}^{2})}$ \citep{ZhangB2004}, which is illustrated in Fig.\ref{fig:cartoon} for $\theta_{\rm c}\sim 7$ deg and $\epsilon_{0}=10^{51}$ erg; (c) the two component jet model \citep{Berger2003,WuXF2005,Huang2006}.
The SGRB outflows are likely structured, too, as found in the numerical simulations \citep{Aloy2005,Murguia-Berthier2017} and in the afterglow modeling \citep{JinZP2007}.

A bright SGRB has a typical isotropic-equivalent kinetic energy of $E_{\rm iso}\sim 10^{51}$ erg \citep{ZhangFW2012,Fong2015}. Such energetic outbursts are detectable for {\it Swift}, Fermi-GBM and GECam detectors\footnote{Fermi-GBM and GECam like detectors are benefited for their very wide field of view \citep{Meegan2009,Xiong2017}, which is very important for catching the GRB/GW association events. {\it Swift} has a sensitivity much higher than Fermi-GBM and comparable to GECam but its field of view is just $\sim 1.4-2.4$ sr. The proposing GECam mission \citep{Xiong2017} is designed to cover $90\%$ of the sky at a sensitivity of $2\times10^{-8}{\rm erg~cm}^2{\rm ~s}^{-1}$ in the energy range of $8~{\rm keV}-2~{\rm MeV}$. Its location accuracy is of $\sim 1$ square degrees. All these properties make GECam a very suitable detector to record the GW-associated GRB signals.}
at a distance of $D\leq 200$ Mpc even when our line of sight is ``slightly" outside of the ejecta  (note that in this paragraph we adopt {\it the uniform jet model}). It is known the on-beam and off-beam isotropic energies are different by a factor of $\sim (1+(\eta\Delta \theta)^{2})^{3}$ \citep[for $\Delta \theta\geq\theta_{\rm j}$; see e.g.][]{Lightman1983}, where $\eta$ is the initial Lorentz factor of the GRB outflow and the viewing angle $\theta_{\rm v}=\theta_{\rm j}+\Delta \theta$. Hence we have
\begin{equation}
\Delta \theta \leq 0.04 [{E_{\rm iso}\over 10^{51}~{\rm erg}}]^{1/6}({\eta\over 100})^{-1}({D\over 200~{\rm Mpc}})^{-1/3}({{\cal F}_{\rm th}\over 10^{-7}~{\rm erg~cm^{-2}}})^{-1/6},
\end{equation}
where ${\cal F}_{\rm th}\sim 10^{-7}~{\rm erg~cm^{-2}}$ is the fluence threshold for a reliable detection of a weak GRB by Fermi-GBM like detector. The $\eta$ has been normalized to $\sim 100$, motivated by the Lorentz factor$-$luminosity correlation of GRBs \citep{LuJ2012,Liang2010,FanYZ2012}. Unless $\eta\leq 50$, we have $\Delta \theta < \theta_{\rm j}(\sim 0.1)$, implying that in the uniform jet model, the off-beam events can enhance the GRB/GW association at most moderately (the enhancement factor is estimated as ${\cal R}\approx (\theta_{\rm j}+\Delta \theta)^{2}/\theta_{\rm j}^{2}$). Such events should be observed mainly in X-rays and last longer, since the energy of the gamma-rays is lowered by a factor of $a_{0}=[1+(\eta\Delta \theta)^{2}]^{-1}$ and the duration is extended by a factor of $a_{0}^{-1}$.

{\it In the structured jet model}, GRB/GW associations are more common. In the numerical simulations, relativistic outflows with $\eta\geq 10$ and $\epsilon(\theta\leq \theta_{\rm cut})\geq 10^{48}$ erg are found within the polar angle of $\theta_{\rm cut} \lesssim 0.3-0.4$ rad \citep{Aloy2005,GOTTLIEB2017,Murguia-Berthier2017}.  The photospheric radius of a relativistic outflow reads $R_{\rm ph}\approx 4\times 10^{11}~{\rm cm}~(L_{\rm tot}/10^{48}~{\rm erg~s^{-1}})(\eta/10)^{-3}$, where $L_{\rm tot}$ is the total luminosity of the outflow \citep{Paczynski1990}. The initial radius of the ``reborn" fireball is $R_0 \sim 10^{9}$ cm \citep{Aloy2005}. As long as $R_{\rm ph}$ is in the same order of $\eta R_0$, the thermal radiation will be efficient, requiring that \citep[one can get the following expression with for example eq.(9) of][by setting $f\sim 1$]{Fan2011}
\begin{equation}
\eta\sim 25 (L_{\rm tot}/10^{48}~{\rm erg~s^{-1}})^{1/4}(R_0/10^{9}~{\rm cm})^{-1/4}.
\end{equation}
The temperature of the emission can be estimated as $T_{\rm obs} \sim 20~{\rm keV}(L_{\rm tot}/10^{48}~{\rm erg~s^{-1}})^{1/4}(R_{0}/10^{9}~{\rm cm})^{-1/2}(\eta/25)^{8/3}$. The corresponding peak energy of the observed spectrum ($\nu f_{\nu}$) is thus
\begin{equation}
E_{\rm p} \sim 3.92T_{\rm obs} \sim 78~{\rm keV}(L_{\rm tot}/10^{48}~{\rm erg~s^{-1}})^{1/4}(R_{0}/10^{9}~{\rm cm})^{-1/2}(\eta/25)^{8/3},
\end{equation}
where the redshift correction has been ignored since in this work we concentrate on the nearby events. The emission duration is likely determined by the width of the ejecta in the direction of the line of sight. Such emission, if within a distance of $\sim 200$ Mpc, are detectable for {\it Swift}, GECam and Fermi-GBM like detectors since the corresponding flux is $\gtrsim 10^{-7}~{\rm erg~cm^{-2}~s^{-1}}$ for a $\gamma-$ray luminosity of $\sim 5\times 10^{47}~{\rm erg~s^{-1}}$.  The association chance between these weak GRB-like transients and GW events is larger than that between the bright SGRBs and GW events by a factor of
\begin{equation}
{\cal R}\sim {1-\cos\theta_{\rm cut} \over 1-\cos \theta_{\rm j}}\sim 16{(\theta_{\rm cut}/0.4~{\rm rad})^2\over (\theta_{\rm j}/0.1~{\rm rad})^{2}},
\end{equation}
implying a more promising prospect of establishing the GRB/GW association in the near future. Note that here $\theta_{\rm cut}\sim 0.4$ rad is adopted to match the structured jet edge found in some numerical simulations. Moreover, a Gaussian-type jet $\epsilon(\theta)\approx 10^{50-51}~{\rm erg}~\exp{(-\theta^2/2\theta_{\rm c}^{2})}$ will yield $\epsilon(\theta=0.38~{\rm rad})\sim 10^{47-48}~{\rm erg}$ for $\theta_{\rm c}\sim 0.1$ rad, which is detectable for the GECam detector as long as the events are within a distance of $\sim 200$ Mpc \citep{Xiong2017}.
The other prediction of the off-axis ejecta model is an unambiguous
re-brightening of the afterglow due to the emergence of the forward shock
emission of the energetic ejecta core \citep{WeiDM2003,Kumar2003}. Recently, \citet{Lamb2017} calculated such emission and suggested them as one of the most promising electromagnetic counterparts of neutron star mergers that may be able to outshine the macronova/kilonova emission (These authors also mentioned the prompt emission but did not go further). Since the beam-corrected SGRB rate (see eq.(\ref{eq:Rnsm})) is roughly comparable to the gravitational wave event based estimate \citep{LVC2017}, it may be reasonable to speculate that SGRBs were produced in a good fraction of mergers, for which the GRB/GW association probability may be as high as $\sim\theta_{\rm cut}^{2}/2\sim 10\%$. The prospect of establishing the GRB/GW association is thus more promising than that suggested in the literature \citep{Williamson2014,Clark2015,LiX2016ApJ}.

Very recently, \citet{Lazzati2017} calculated the emission from the possible wide cocoon (i.e., within the polar angle $\sim 40^{\circ}$) surrounding the SGRB outflow and suggested the  prompt X-ray emission with $E_{\rm iso}\sim 10^{49}~{\rm erg}$, which is significantly stronger than our signal. Such kind of energetic shortly-lasting X-ray outbursts will be nice electromagnetic counterparts of GW events \citep[see however][for the results based on the 3 dimensional simulation]{GOTTLIEB2017}.

In the above discussion, the successful launching of GRB ejecta is assumed. This may be not always the case. As already revealed by the numerical simulations, some relativistic ejecta can not break out successfully if the initial half-opening angles are too wide \citep[see e.g.][]{Aloy2005,Nagakura2014,Murguia-Berthier2017}. In such cases, low luminosity events may be powered by the mildly-relativistic outflow. The simple thermal radiation model, however, is usually unable to give rise to significant emission at energies above 10 keV for $\eta\leq 10$ (not that as long as $R_{\rm ph}\gg \eta R_0$, we have $E_{\rm p}\propto \eta^{8/3}$) and additional physical process(es) (e.g., shocks, or magnetic energy dissipation, or shock breakout) should be introduced to generate X-ray/gamma-ray emission.

\section {Summary and discussion}

After the discovery of the SGRB afterglow, dedicated efforts have been made to identify the jet breaks and then infer the half-opening angles. However, the sample increases rather slowly due to the dim nature of these events.
The main reason for the non-detection/identification of the jet breaks in most SGRB afterglows may be the lack of deep follow-up observations.
The realization that the macronovae may appear within one to two weeks after the GRBs inspired the very late afterglow observations. With these high-quality optical/near-infrared data we found two jet breaks in GRB 150424A and GRB 160821B. Together with the previous results, we have a SGRB/jet sample consisting of ten events (including one long-short event). The inferred half-opening angles have a very narrow distribution (i.e., $\theta_{\rm j}\sim 0.1$). Though the sample is still small, there are four events taking place locally (i.e., $z\leq 0.2$), with which the ``local" neutron star merger rate density has been estimated {to be $\sim 1109~{\rm Gpc^{-3}~yr^{-1}}$ or $\sim 162~{\rm Gpc^{-3}~yr^{-1}}$ if the narrowly-collimated GRB 061201 is excluded.} These SGRB-based local neutron star merger rates, however, are conservative since a few ``local" SGRBs (including GRB 080905A and GRB 150101B) have not been taken into account, and moreover just $\sim 1/4$ SGRBs have redshifts. Further enhancement is plausible if the SGRB production fraction of neutron star mergers is lower than $100\%$ and the SGRB detection rate by {\it Swift} seems to be lower than BATSE-like detectors. {Nevertheless, more local SGRBs with reasonably measured $\theta_{\rm j}$ are needed to get more reliable ${\cal R}_{\rm nsm}$ since the current estimate is seriously affected by the narrowly-beamed event GRB 061201.}

We have also examined the HST data of GRB 150424A and GRB 160821B  to search for possible macronova signal(s). In GRB 150424A no sign has been found. While in GRB 160821B, the HST and Keck \citep{Kasliwal2017a} data at $t\sim 3.6-4.3$ days can be interpreted as a power-law afterglow component plus a thermal component with a temperature of $\sim 3100$ K. However, with the currently rather-limited (publicly-)available data, no evidence as strong as that for GRB 130603B, GRB 060614, GRB 050709 and GRB 170817A can be provided.

Finally, motivated by the plausible promising detection prospect of neutron star mergers in the near future, we have re-estimated the GRB/GW association probability. For the very nearby (i.e., $D\leq 200$ Mpc) events, some off-beam GRBs (in the uniform jet model) may be detectable, possibly appearing as the low-luminosity X-rich transients/GRBs (if the duration of the intrinsic (on-beam) event is dominated by one single pulse, the observed duration would be extended by a factor of $1+(\eta \Delta \theta)^{2}$).
The enhancement of the GRB/GW association is, however, at most moderate. The situation is different if the merger-driven relativistic ejecta are structured in a wide solid angle. In such a case, the prompt emission of the relativistic ejecta, though viewed off-axis, are detectable for {\it Swift}, Fermi-GBM and GECam like detectors if the sources are at $D\leq 200$ Mpc and the corresponding GRB/GW association probability may be high up to $\sim 10\%$.

\section*{Acknowledgments}
We thank the anonymous referee for helpful suggestions and T. Piran, R. F. Shen and Y. M. Hu for discussions.
This work was supported in part by 973 Programme of China (No. 2014CB845800), by NSFC under grants 11525313 (the National Natural Fund for Distinguished Young Scholars), 11433009 and 11773078, by the Chinese Academy of Sciences via the Strategic Priority Research Program (No. XDB23040000), Key Research Program of Frontier Sciences (No. QYZDJ-SSW-SYS024) and the External Cooperation Program of BIC (No. 114332KYSB20160007).
\\

\clearpage

\end{document}